\documentclass[11pt,a4paper]{article}
\usepackage{jheppub}

\usepackage[utf8]{inputenc}
\usepackage{amsmath,amssymb}
\usepackage{comment}
\usepackage{amsmath}
\usepackage{braket}
\usepackage{mathrsfs}
\usepackage{mathtools}
\usepackage{array}
\usepackage{fancyhdr}
\usepackage{color}
\usepackage{bm}
\usepackage{url}
\usepackage{multirow}
\usepackage{float}
\usepackage{makecell}

\usepackage{ragged2e}
\usepackage[justification=justified]{caption} 

\usepackage{diagbox}
\usepackage{nicematrix}
\usepackage{tikz}
\usetikzlibrary{patterns}
\usetikzlibrary{arrows.meta}

\allowdisplaybreaks

\definecolor{RedWine}{rgb}{0.743,0,0}
\definecolor{RoyalBlue}{rgb}{0.25,.41,.88}
\definecolor{ForestGreen}{rgb}{.13,.54,.13}

\usepackage{tensor}
\usepackage{xcolor}
\usepackage{slashed}
\usepackage{tikz}
\usepackage{dsfont} 
\usepackage{bbold}

\renewcommand{\thepage}{\arabic{page}}
\begin{document}

\title{Bosonization for the Schwinger effect at small fermion mass: analytic formula for the 1+1D current in an arbitrary time-dependent, inhomogeneous electric field}

\author[a]{Samuel E. Gralla}
\author[a]{and Morifumi Mizuno}

\affiliation[a]{Department of Physics, University of Arizona, Tucson, Arizona 85721, USA}

\abstract{We derive an analytic formula for the expectation value of the current due to Schwinger pair-production in an arbitrary time-dependent, spatially inhomogeneous external electric field in 1+1 dimensions, as a perturbative expansion in the fermion mass $m$.  
We work to order $m^2$, finding contributions at order $m^0$, $m^2 \log m$ and $m^2$; there is no contribution at $O(m)$.  The calculation is enabled by a new ``hybrid method'' using bosonization, where part of the calculation is performed on the fermion side and part is performed on the boson side.  The fermion side allows us to define the vacuum using the free theory in the infinite past, while the boson side allows us to compute the current without regularization, since the current operator is linear in the boson field. 
In the special case of a spatially homogeneous background field, we also perform the calculation purely on the fermion side using adiabatic regularization.  The detailed agreement of the results provides a highly non-trivial check of the duality.}  

\maketitle

\section{Introduction}

The Schwinger effect is the instability of the vacuum to the production of pairs of electrons and positrons in the presence of an external electric field \cite{Heisenberg:1936nmg,Schwinger1951jun,Brezin.Itzykson1970oct,Cohen.McGady2008aug}.  Beyond its role as a benchmark experimental prediction, it provides a helpful analog to strong-gravity physics, sharing many mathematical features with the theory of particle creation in curved spacetime \cite{Birrell:1982ix,Fulling1989,Parker:2009uva}.  The effect is most commonly studied using a uniform electric field, which clearly reveals its basic features.  However, this idealization of an eternally existing, space-filling source also complicates the physical interpretation, especially with regard to Lorentz-invariance \cite{Garriga:2012qp}.

The connection to experiment becomes more direct if Lorentz-invariance is broken by a prescribed evolution of the electric field.  An experimenter who turns on and off an electric field over a finite region of space will naturally prescribe a time-dependent, spatially inhomogeneous background field.   However, it is less clear what can actually be measured in this context, since the notion of particle becomes ambiguous during the evolution of the electric field \cite{Dabrowski.Dunne2016Sep,Ilderton2022jan}.  One way to sidestep these difficulties is to instead consider the \textit{current} produced by the particles, which corresponds to a local operator whose expectation value could in principle be measured by repeating the experiment many times.  Unfortunately, this adds significant difficulty to the computation, since one must regularize infinities arising from the coincidence limit of the operators defining the current.  While there has been much progress with adiabatic methods \cite{Parker.Fulling1974jan,Parker:2009uva} or point-splitting \cite{DeWitt1975aug,Christensen1976nov,Christensen1978feb,Herman.Hiscock1996mar} in both electromagnetic and gravitational settings \cite{Anderson1990,Bunch.Davies1997jana,Candelas1980,Frolov1989,Hayashinaka.Yokoyama2016jul,Howard1984,Kobayashi.Afshordi2014oct,Stahl2016,DelRio2014,Landete2013,Ferreiro.Navarro-Salas2018jun,Beltran-Palau.Navarro-Salas.ea2020may}, the fully generic case of time-dependent, spatially inhomogeneous fields remains unsolved.

In this paper we will show that in the simplified setting of 1+1 dimensions and a small fermion mass $m$,\footnote{Physically, the small-mass expansion restricts us to experiments with duration less than the timescale $1/m$ set by the fermion---see further discussion in Sec.~\ref{sec: validity of m expansion} below.} the need for regularization can be sidestepped entirely using the method of bosonization \cite{Coleman1975apra,Mandelstam1975maya,Haldane:1981zza,Witten:1983ar,Naon1985apr,vonDelft:1998pk,Senechal:1999us,Shankar:2017zag,Justine2021}.  The basic point is that the current operator becomes \textit{linear} in the boson field, so there are no divergences resulting from products of operators.  Our expansion in the mass $m$ allows us to compute on the boson side, treating the sine-Gordon term as an interaction.  However, a naive perturbation expansion encounters divergences, which we attribute to the fact that the correct vacuum state (corresponding to the fermion vacuum) is not smoothly related to the boson Fock vacuum \cite{Lukyanov.Zamolodchikov1997jun}.  Fortunately, we find that the vacuum state issues in effect decouple from the effects of the background field, allowing us to handle the vacuum-sensitive parts of the calculation on the fermion side (with a free fermion), while the background-sensitive parts are done on the boson side (where no regularization is needed).   Using this hybrid boson-fermion perturbation method, we obtain a rather remarkable result (Eq.~\eqref{eq:j expectation value boson smallm} below): an explicit analytic formula for the expectation value of the current in the presence of an  \textit{arbitrary time-dependent, spatially homogeneous electric field} to second order in the fermion mass.

Deriving this result purely on the fermion side would require solving the Dirac equation in an arbitrary background field and regularizing the coincidence limit to obtain a finite current.  While the full calculation remains out of reach, a practical regularization method has been developed for spatially homogeneous (but time-dependent) fields in \cite{Ferreiro.Navarro-Salas2018jun}.  We perform this procedure perturbatively in $m$ and compare with the results of the boson-fermion hybrid method.  We find perfect agreement, providing a highly non-trivial check of the duality, including the role of regularization of composite operators.

This paper is organized as follows. In section \ref{sec:boson review}, we review bosonization of the free fermion in an external field.
In section \ref{sec:expectation value of current}, we explain the perturbative expansion and apply the hybrid method to derive the current expectation value to second order in the fermion mass. 
In section \ref{sec:fermi Schwinger}, we compute the same quantity on the fermion side in the special case of spatially homogeneous fields and verify that the results agree. In section \ref{sec:discussion}, we discuss some implications and potential extensions of our results.  Various technical details are deferred to four appendices.  We use units with take $c=\hbar=1$, and the signature of the metric is $(-,+)$.

\section{Bosonization of fermions in background fields}\label{sec:boson review}
In this section, we review the process of bosonization and apply it to 1+1 dimensional QED with an external field in the no-backreaction approximation. Our discussion closely follows our paper \cite{Gralla:2025imj} (hereafter paper I), in which supporting detail may be found.  We take the gamma matrices in the following representation: 
\begin{align}
    \label{eq:gamma matrix}
    \gamma^{t}=\begin{pmatrix}
        0&1\\1 &0
    \end{pmatrix},
    \quad 
    \gamma^{x}=\begin{pmatrix}
        0&1\\-1 &0
    \end{pmatrix},
     \quad
     \gamma^{5}\equiv\gamma^{t}\gamma^{x}=\begin{pmatrix}
        -1&0\\0 &1
    \end{pmatrix}.
\end{align}
We work in the Schr{\"o}dinger picture with Coulomb gauge for both the quantum field $\hat{A}^{\mu}=(\hat{A},0)$ and external classical field $A_C^{\mu}=(A_C,0)$. Paper I reviews the process of eliminating $\hat{A}$ by solving the constraint equation, leaving
\begin{align}
    \label{eq: QED Hamiltonian}
    \widehat{H}_{\rm QED}(t)=&\int dx:\left[-i\overline{\widehat{\psi}}\gamma^{x}\partial_{x}\widehat{\psi}+m\overline{\widehat{\psi}}\widehat{\psi}+qA_{C}(t,x)\widehat{\psi}^{\dagger}\widehat{\psi}\right]:\notag
    \\
    &\quad-\frac{q^{2}}{4}\iint dxdy:\!\widehat{\psi}^{\dagger}(x)\widehat{\psi}(x)\!:|x-y|:\!\widehat{\psi}^{\dagger}(y)\widehat{\psi}(y)\!:,
\end{align}
where $\overline{\widehat{\psi}}(x)=\widehat{\psi}^{\dagger}(x)\gamma^t$ and the spatial integral is taken from $-\infty$ to $\infty$.  The symbol $:\cdots:$ denotes normal-ordering of fermion operators.  In this paper we will have to be quite careful about the meaning of normal-ordering, since it will be performed on both sides of the duality, and with respect to different choices of mass.  However,  at this stage we leave the mass arbitrary, since a change of normal-ordering mass merely shifts the Hamiltonian \eqref{eq: QED Hamiltonian} by a  $c$-number (Appendix~\ref{app:fermi normal order}). 
The fermion operators satisfy the usual anti-commutation relations:
\begin{align}
    \label{eq:fermi anti commutation1}
    \left\{\widehat{\psi}_{\alpha}(x),\widehat{\psi}^{\dagger}_{\beta}(y)\right\}
    =&\delta_{\alpha\beta}\delta(x-y),
    \\
    \label{eq:fermi anti commutation2}
     \left\{\widehat{\psi}_{\alpha}(x),\widehat{\psi}_{\beta}(y)\right\}
    =&0.
\end{align}
The Coulomb interaction is encoded in the last term of \eqref{eq: QED Hamiltonian}.  We will neglect this term, which corresponds to neglecting the electromagnetic field of newly created particles (no backreaction). 
A mathematically controlled version of the no-backreaction condition is to take the limit $q\to0$ while holding $qA_{C}$ (and equivalently $qE_C$) fixed. Thus, $E_C$ scales as $\sim1/q$ and as $q\to0$, $E_C\gg q$. Since $q$ and $E$ have the same dimension in 1+1D,\footnote{Gauss's law in 1+1D shows that $E_C$ and $q$ have the same dimension.  Since $qE_c$ has the dimension of force (mass dimension of 2), $q$ and $E_C$ both have the mass dimension.} $q$ sets the typical scale of the electric field created by charged particles; thus, $E_C\gg q$ implies that the field created by particles is much smaller than the external field. 
Then, in this limit, the Hamiltonian of the system becomes 
\begin{align}\label{eq: fermion quantum hamiltonian}
    \widehat{H}(t)=&\int dx:\left[-i\overline{\widehat{\psi}}\gamma^{x}\partial_{x}\widehat{\psi}+m\overline{\widehat{\psi}}\widehat{\psi}+qA_{C}(t,x)\widehat{\psi}^{\dagger}\widehat{\psi}\right]:.
\end{align}  
Now, we bosonize this system (e.g. \cite{Coleman1975apra,Mandelstam1975maya,Haldane:1981zza,Witten:1983ar,Naon1985apr,vonDelft:1998pk,Senechal:1999us,Shankar:2017zag,Justine2021}). 
To do so, we accept simple substitution rules to obtain the boson theory (e.g. \cite{Coleman.Jackiw.ea1975sep,Coleman1976sep,Smilga:1992hx}):
\begin{align}
    \label{eq:fermi=boson kinetic}
    :-i\widehat{\psi}(x)\gamma^{x}\partial_{x}\widehat{\psi}(x):
    &\longleftrightarrow N_{\mu}\left\{\frac{1}{2}\widehat{\Pi}^{2}(x)+\frac{1}{2}(\partial_{x}\widehat{\phi}(x))^{2}\right\},
    \\
    \label{eq:fermi=boson jt}
    :\overline{\widehat{\psi}}(x)\gamma^{t}\widehat{\psi}(x):&\longleftrightarrow \frac{1}{\sqrt{\pi}}\partial_{x}\widehat{\phi}(x),
    \\
    \label{eq:fermi=boson jx}
    :\overline{\widehat{\psi}}(x)\gamma^{x}\widehat{\psi}(x):&\longleftrightarrow-\frac{1}{\sqrt{\pi}}\widehat{\Pi}(x),
    \\
     \label{eq:fermi=boson mass}
    :\overline{\widehat{\psi}}(x)\widehat{\psi}(x):&\longleftrightarrow-\frac{e^{\gamma}\mu}{2\pi}N_{\mu}\left\{\cos{\sqrt{4\pi}\widehat{\phi}(x)}\right\},
    \\
     \label{eq:fermi=boson chiral}
    i:\overline{\widehat{\psi}}(x)\gamma^{5}\widehat{\psi}(x):&\longleftrightarrow\phantom{a}\frac{e^{\gamma}\mu}{2\pi}N_{\mu}\left\{\sin{\sqrt{4\pi}\widehat{\phi}(x)}\right\}.
\end{align}
The symbol $N_{\mu}\left\{\cdots\right\}$ denotes  boson normal-ordering with respect to an  arbitrary mass $\mu$.  That is, if we define the creation and annihilation operators $\widehat{a}^{\dagger}_{k,\mu}$ and $\widehat{a}_{k,\mu}$ by mode expanding $\widehat{\phi}(x)$ and $\widehat{\Pi}(x)$ as 
\begin{align}
    \label{eq:phi expansion}
    \widehat{\phi}(x)&=\int_{-\infty}^{\infty}\frac{dk}{2\pi\sqrt{2\omega_{k,\mu}}}\left(\widehat{a}_{k,\mu}e^{ikx}+\widehat{a}^{\dagger}_{k,\mu}e^{-ikx}\right),
    \\
    \label{eq:Pi expansion}
    \widehat{\Pi}(x)&=
    -i\int_{-\infty}^{\infty}\frac{dk}{2\pi}\sqrt{\frac{\omega_{k,\mu}}{2}}
    \left(\widehat{a}_{k,\mu}e^{ikx}-\widehat{a}^{\dagger}_{k,\mu}e^{-ikx}\right),
\end{align}
with
\begin{align}\label{eq:frequency}
    \omega_{k,\mu}=\sqrt{k^{2}+\mu^{2}},
\end{align}
then $N_{\mu}\left\{\cdots\right\}$ means to rearrange the order of operators by placing all $\widehat{a}_{k,\mu}$ to the right of all $\widehat{a}^{\dagger}_{k,\mu}$. The Boson field operators satisfy the canonical commutation relations
\begin{align}
    \label{eq:commu phi,pi}
    \left[\widehat{\phi}(x),\widehat{\Pi}(y)\right]=&i\delta(x-y),
    \\
    \label{eq:commu phi,phi}
    \left[\widehat{\phi}(x),\widehat{\phi}(y)\right]=&0,
    \\
    \label{eq:commu pi,pi}
    \left[\widehat{\Pi}(x),\widehat{\Pi}(y)\right]=&0,
\end{align}
meaning that $\widehat{a}_{k,\mu}$ and $\widehat{a}^{\dagger}_{k,\mu}$ satisfy the usual commutation relations
\begin{align}
    \label{eq:commutation a-adagger}
    \left[\widehat{a}_{k,\mu},\widehat{a}^{\dagger}_{k',\mu}\right]=&2\pi\delta(k-k'),
    \\
    \left[\widehat{a}_{k,\mu},\widehat{a}_{k',\,\mu}\right]=&0,
    \\
    \left[\widehat{a}^{\dagger}_{k,\mu},\widehat{a}^{\dagger}_{k',\mu}\right]=&0. \label{eq:commutation adagger-adagger}
\end{align}
An important fact is that the operator $\mu N_\mu$ is actually independent of $\mu$ when acting on an exponential operators $e^{\pm i \sqrt{4\pi} \widehat{\phi}(x)}$ \cite{Coleman1975apra}.  Thus, the right-hand sides of \eqref{eq:fermi=boson mass} and \eqref{eq:fermi=boson chiral} are actually independent of the arbitrary mass $\mu$.

The arrows used in the substitution rules~\eqref{eq:fermi=boson kinetic}--\eqref{eq:fermi=boson chiral} have slightly different meanings depending on the operator. 
For Eq.~\eqref{eq:fermi=boson kinetic}, changing either the fermion normal-ordering mass or the boson normal-ordering mass $\mu$ shifts the corresponding side by a constant $c$-number. Thus, the arrow in Eq.~\eqref{eq:fermi=boson kinetic} should be understood as an equivalence only at the level of the Hamiltonian, where such a shift does not affect the dynamics. 
For the operators~\eqref{eq:fermi=boson jt}, \eqref{eq:fermi=boson jx}, and \eqref{eq:fermi=boson chiral}, the normal-ordering on the LHS is independent of the choice of the normal-ordering mass (see Appendix~\ref{app:fermi normal order}), while no normal-ordering ambiguity is present on the boson side. Therefore, we regard the arrows in Eqs.~\eqref{eq:fermi=boson jt}, \eqref{eq:fermi=boson jx}, and \eqref{eq:fermi=boson chiral} as exact equalities.
For Eq.~\eqref{eq:fermi=boson mass}, changing the fermion normal-ordering mass shifts the LHS by a $c$-number, whereas the RHS is independent of $\mu$. Thus, substitution at the level of the Hamiltonian is valid for any choice of fermion normal-ordering mass. However, as we will show later, the hybrid method agrees with the pure fermion calculations precisely by interpreting the arrow in Eq.~\eqref{eq:fermi=boson mass} as an exact equality that holds when the fermion normal-ordering mass is set to be zero.  This is consistent with Coleman's original derivation of the duality using massless normal-ordering \cite{Coleman1975apra}.  We therefore use Eqs.~\eqref{eq:fermi=boson mass} as an exact equality only with the zero-mass choice of fermion normal-ordering.

With these observations in mind, after the substitution \eqref{eq:fermi=boson kinetic}--\eqref{eq:fermi=boson chiral} into Eq.~\eqref{eq: fermion quantum hamiltonian}, we obtain the bosonized version of the same theory as 
\begin{align}
    \label{eq: boson quantum Hamiltonian}
    \widehat{H}(t)=\int dxN_{\mu}\left\{\frac{1}{2}\widehat{\Pi}(x)^{2}+\frac{1}{2}\left(\partial_{x}\widehat{\phi}(x)\right)^{2}
    +\frac{q}{\sqrt{\pi}}E_{C}(t,x)\widehat{\phi}(x)
    -\frac{e^{\gamma}m\mu}{2\pi}\cos{\sqrt{4\pi}\widehat{\phi}(x)}
    \right\}
\end{align}
where $E_{C}$ is given by
\begin{align}
    E_C=-\partial_x A_C.
\end{align}
The physical content of Eq.~\eqref{eq: boson quantum Hamiltonian} is independent of the normal-ordering mass $\mu$.  For the first three terms of Eq.~\eqref{eq: boson quantum Hamiltonian}, the change of the normal-order mass $\mu$ merely results in shift of the Hamiltonian by a $c$-number, while the combination $\mu N_{\mu}\{\cos{\sqrt{4\pi}\widehat{\phi}(x)}\}$ is $\mu$-independent \cite{Coleman1975apra}. 

\section{Hybrid boson-fermion calculation for a general background field}\label{sec:expectation value of current}
Our goal in this paper is to calculate the expected value of current in response to an imposed electric field.  We assume that the initial state is vacuum with zero electric field.  The electric field is turned on at time $t=t_0$ and allowed to have arbitrary time and space dependence thereafter.

\subsection{Setup}
The vacuum state $\ket{\Omega}$ is defined as the lowest-energy eigenvector of the Hamiltonian for $t<t_0$ (before the electric field is turned on).  This definition can in principle be applied on either side of the duality, but the fermion side is more tractable.  We expand the fermion operator $\widehat{\psi}(x)$ using the physical fermion mass $m$ as
\begin{align}
    \label{eq:psi define b and d}
    \widehat{\psi}(x)=\begin{pmatrix}
        \widehat{\psi}_{1}(x)\\
        \widehat{\psi}_{2}(x)
    \end{pmatrix}
    =&\int_{-\infty}^{\infty}\frac{dk}{2\pi\sqrt{2\omega_{k,m}}}
    \left[
    \widehat{b}_{k}e^{ikx}\begin{pmatrix}
        \sqrt{\omega_{k,m}-k}\\
        \sqrt{\omega_{k,m}+k}
    \end{pmatrix}
    +\widehat{d}^{\dagger}_{k}e^{-ikx}\begin{pmatrix}
        -\sqrt{\omega_{k,m}-k}\\
        \sqrt{\omega_{k,m}+k}
    \end{pmatrix}
    \right],
\end{align}
where $\omega_{k,m}$ is defined by Eq.~(\ref{eq:frequency}). The canonical anti-commutation relation (\ref{eq:fermi anti commutation1}) and (\ref{eq:fermi anti commutation2}) implies that $\widehat{b}_{k}$ and $\widehat{d}_{k}$ satisfy the standard anti-commutation relations:
\begin{align}
    \left\{\widehat{b}_{k},\widehat{b}^{\dagger}_{k'}\right\}=  \left\{\widehat{d}_{k},\widehat{d}^{\dagger}_{k'}\right\}=2\pi\delta(k-k'),\label{eq:fermicommutation}
\end{align}
and the rest are all zero. We can then define the usual Fock vacuum $\ket{0_{\rm f},m}$ by
\begin{align}\label{eq: fermion vacuum}
    \widehat{b}_{k}\ket{0_{\rm f},m}=\widehat{d}_{k}\ket{0_{\rm f},m}=0
\end{align}
for all $k$. Since the Hamiltonian (\ref{eq: fermion quantum hamiltonian}) at $t<t_{0}$ is just the free-fermion Hamiltonian, Eq.~(\ref{eq: fermion vacuum}) is in fact the lowest energy state for $t<t_{0}$.  That is, we have
\begin{align}
    \label{eq: fermi vacuum = boson vacuum}
    \ket{\Omega}=\ket{0_{\rm f},m}.
\end{align}
This relates the fermion Fock vacuum to the true vacuum.  There is no correspondingly simple expression on the boson side of the theory, which is fully interacting at finite $m$.

We take the system to be in the vacuum state initially, but it evolves away from this state after the field is turned on at $t=t_0$.  If $\widehat{U}$ denotes time-evolution operator determined by the Hamiltonian $\widehat{H}$, the state $\ket{\psi(t)}$ evolves as
\begin{align}\label{eq:verysimple}
    \ket{\psi(t)} \simeq \widehat{U}(t,t_0) \ket{\Omega},
\end{align}
where $\simeq$ denotes equality up to phase. The electric current operator is given by the LHS of Eq.~\eqref{eq:fermi=boson jx} multiplied by the charge $q$. Since we treat the arrow for Eq.~\eqref{eq:fermi=boson jx} as equality (see discussion below Eq.~\eqref{eq:commutation adagger-adagger}), the vacuum expectation value $ \braket{j^{x}}(t,x)$ may be written as
\begin{align}
    \label{eq:<j> definition}
    \braket{j^{x}}(t,x)&=-\frac{q}{\sqrt{\pi}}\bra{\psi(t)}\widehat{\Pi}(x)\ket{\psi(t)}.
\end{align}
Eq.~(\ref{eq:<j> definition}) shows that there is no evaluation of the coincidence limit of composite operators, which is an advantage of working on the boson side.

\subsection{Interaction picture}
Our goal is to expand in the mass $m$.  It is therefore natural to separate the Hamiltonian (\ref{eq: boson quantum Hamiltonian}) into a free part and a cosine interaction.  However, this division leads to to difficulties because the free part is massless, whereas its operators $\widehat{\phi}(x)$ and $\widehat{\Pi}(x)$ are expanded in mass-$\mu$ creation and annihilation operators as Eqs.~(\ref{eq:phi expansion}) and (\ref{eq:Pi expansion}).  We cannot simply set $\mu=0$ because this mass is required to define the interaction term.  Instead we follow Coleman \cite{Coleman1975apra} and introduce a fictitious mass term $\mu^{2}\widehat{\phi}^{2}(x)/2$ to the Hamiltonian (\ref{eq: boson quantum Hamiltonian}).  This deformed version of the theory is not the same as Eq.~(\ref{eq: boson quantum Hamiltonian}), but in the limit of $\mu\to0$ where the effect of the fictitious mass becomes absent, it reduces to the original theory; thus, at the end of the calculations, we set $\mu=0$ to recover the physical content of Eq.~(\ref{eq: boson quantum Hamiltonian}). 

We denote the deformed Hamiltonian by $\widehat{H}_{\mu}(t)$ and split it into background and interaction terms as
\begin{align}
    \label{eq:total Hamiltonian}
    \widehat{H}_{\mu}(t)=&\widehat{H}_{0}(t)+\widehat{H}_{\rm int}(t)
\end{align}
where the background itself is split into as
\begin{align}
    \label{eq: define H0}
    \widehat{H}_{0}(t)=\widehat{H}_{\rm  ff}+\widehat{H}_{E}(t).
\end{align}
We defined the ``free-free" Hamiltonian $\widehat{H}_{\rm  ff}$ by
\begin{align}
     \label{eq:free-free Hamiltonian}
    \widehat{H}_{\rm ff}&=\int \! dxN_{\mu}\left\{
    \frac{1}{2}\widehat{\Pi}(x)^{2}+\frac{1}{2}(\partial_{x}\widehat{\phi}(x))^{2}
    +\frac{1}{2}\mu^{2}\widehat{\phi}(x)^{2}
    \right\}=\int\frac{dk}{2\pi}\omega_{k,\mu}\widehat{a}^{\dagger}_{k,\mu}\widehat{a}_{k,\mu},
\end{align}
and $ \widehat{H}_{E}(t)$ and $\widehat{H}_{\rm int}(t)$ are defined by
\begin{align}
    \label{eq:E Hamiltonian}
    \widehat{H}_{E}(t)&=\int\! dx
    \frac{q}{\sqrt{\pi}}E(t,x)\widehat{\phi}(x),
    \\
    \label{eq:int Hamiltonian}
    \widehat{H}_{\rm int}(t)&=\int \!dx\left(-\frac{e^{\gamma}mf(t)}{2\pi}\right)\mu N_{\mu}\left\{
    \cos{\sqrt{4\pi}\widehat{\phi}(x)}
    \right\}.
\end{align}
The role of $f(t)$ here is the same as that in Paper I: it is chosen to vanish at an
early time and to reach $f = 1$ adiabatically at $t = t_0$. Here, adiabatic means the limit where the time derivative of $f(t)$ approaches $0$, i.e., the adiabatic limit ensures that $f(t)\to1$ for fixed $t$. 
After the calculations, we also need to take the $\mu\to0$ limit.

\subsubsection{Vacuum state}

We now express the vacuum state $\ket{\Omega}$ in terms of interaction-picture operators and states. 
In the infinite past, there is no external electric field, and the interaction term is turned off due to $f(t)$.  Thus, the vacuum state of the deformed Hamiltonian (\ref{eq:total Hamiltonian}) at early times is the usual Fock vacuum $\ket{0,\mu}$ satisfying 
\begin{align}
    \widehat{a}_{k,\mu}\ket{0,\mu}=0
\end{align}
for all $k$.  The true vacuum $\ket{\Omega}$ can be approximated by first evolving $\ket{0,\mu}$ from the infinite past to time $t_{0}$ by the deformed Hamiltonian (\ref{eq:total Hamiltonian}) and then taking $\mu\to0$ limit. 
Using $\widehat{U}_{\mu}(t,t')$, $\widehat{U}_{0}(t,t')$, and $\widehat{U}_{\rm int}(t,t')$ defined by Eqs.~\eqref{eq:define U(t,t')}, \eqref{eq:define U0(t,t')}, and \eqref{eq:define Uint(t,t')}, we have
\begin{align}
    \label{eq:sG vacuum}
    \ket{\Omega}&\simeq \lim_{\mu\to0}\widehat{U}_{\mu}(t_{0},-\infty)\ket{0,\mu}
    \\
    &= \lim_{\mu\to0}\widehat{U}_{\rm int}(t_{0},-\infty)\widehat{U}_{0}^{\dagger}(-\infty,t_{0})\ket{0,\mu}
\end{align}
In the first line we use the adiabatic approximation, justified by a slowly-varying $f(t)$.  In the second line we used  Eq.~\eqref{eq:define Uint(t,t')} with $\widehat{U}_T=\widehat{U}_{\mu}$.  Within the time range $(-\infty,t_{0}]$, the external field $E_{C}(t,x)$ is zero, and thus $\widehat{U}_{0}(t_{0},-\infty)$ only changes $\ket{0,\mu}$ by a phase, 
\begin{align}
    \widehat{U}_{0}^{\dagger}(-\infty,t_{0})\ket{0,\mu}\simeq \ket{0,\mu}.
\end{align}
Thus we have
\begin{align}
\ket{\Omega}
    &\simeq \lim_{\mu\to0} \widehat{U}_{\rm int}(t_{0},-\infty)\ket{0,\mu}.\label{eq:sG vacuum int}
\end{align}
Note that the limit of $\mu\to0$ is understood as being taken at the $c$-number level (i.e., after $c$-number quantities, such as expectation values, are obtained). 

\subsubsection{Operators}
In this section we show that the interaction picture operators $\widehat{\phi}^{I}(t,x)$ and $\widehat{\Pi}^{I}(t,x)$ split into ``classical'' and quantum parts in a nice way and derive a useful expression for the expectation value of normal-ordered operators in the Fock vacuum $\ket{0,\mu}$.

Interaction-picture operators evolve according to the background Hamiltonian, as given in Eq~.\eqref{eq:define interaction O} or equivalently Eq.~\eqref{eq:heisenberg O interaction}.  We begin with the creation operator $\widehat{a}^{I}_{k,\mu}(t)$, 
whose time derivative \eqref{eq:heisenberg O interaction} is
\begin{align}
    \label{eq: dynamical a(t)}
    \frac{d}{dt} \widehat{a}^{I}_{k,\mu}(t)&=
    i \left[\widehat{H}_{0}^{I}(t),\widehat{a}^{I}_{k,\mu}(t)\right] 
   \notag
    \\
    &=i\widehat{U}_{0}^{\dagger}(t,t_{0}) \left[\widehat{H}_{0}(t),\widehat{a}_{k,\mu}\right] \widehat{U}_{0}(t,t_{0})\notag
    \\
    &=-i\omega_{k,\mu}\widehat{a}^{I}_{k,\mu}(t)-i\frac{q}{\sqrt{\pi}}\frac{1}{\sqrt{2\omega_{k,\mu}}}\int dx e^{-ikx}E_{C}(t,x),
\end{align}
where $\omega_{k,\mu}$ is defined by Eq.~(\ref{eq:frequency}).
From the second line to the third, we have used 
\begin{align}
    \left[\widehat{H}_{0}(t),\widehat{a}_{k,\mu}\right]=-\omega_{k,\mu}\widehat{a}_{k,\mu}-\frac{q}{\sqrt{\pi}}\frac{1}{\sqrt{2\omega_{k,\mu}}}\int dx e^{-ikx}E_{C}(t,x),
\end{align}
which can be explicitly confirmed by directly computing the LHS using Eqs.~\eqref{eq:phi expansion}, \eqref{eq:Pi expansion}, \eqref{eq: define H0}--\eqref{eq:E Hamiltonian} together with the commutation relation of $\widehat{a}_{k,\mu}$ and $\widehat{a}^{\dagger}_{k,\mu}$ \eqref{eq:commutation a-adagger}--\eqref{eq:commutation adagger-adagger}.
The unique solution to Eq.~\eqref{eq: dynamical a(t)} satisfying the condition $\widehat{a}^{I}_{k,\mu}(t_{0})=\widehat{a}_{k,\mu}$ is given by 
\begin{align}
    \label{eq: a interaction}
    \widehat{a}^{I}_{k,\mu}(t)=\widehat{a}_{k,\mu}e^{-i\omega_{k,\mu}(t-t_{0})}-i\frac{q}{\sqrt{\pi}}\frac{1}{\sqrt{2\omega_{k,\mu}}}\int_{t_{0}}^{t}ds\int dxe^{-i\omega_{k,\mu}(t-s)-ikx}E_{C}(s,x).
\end{align}
To obtain $\widehat{a}^{I\dagger}_{k,\mu}(t)$, one needs only take the Hermitian conjugate of Eq.~\eqref{eq: a interaction}. 

The interaction-picture operators  $\widehat{\phi}^{I}(t,x)$ and $\widehat{\Pi}^{I}(t,x)$ can now be readily obtained.  By sandwiching $\widehat{\phi}(x)$ and $\widehat{\Pi}(x)$ (given by Eqs.~\eqref{eq:phi expansion} and \eqref{eq:Pi expansion}) with $\widehat{U}_{0}^{\dagger}(t,t_0)$ and $\widehat{U}_{0}(t,t_0)$, we can write
\begin{align}
    \label{eq:op sol phi interaction intermidiate}
     \widehat{\phi}^{I}(t,x)&= \int_{-\infty}^{\infty}\frac{dk}{2\pi\sqrt{2\omega_{k,\mu}}}\left(\widehat{a}^{I}_{k,\mu}(t)e^{ikx}+\widehat{a}^{I \dagger}_{k,\mu}(t)e^{-ikx}\right),
     \\
     \label{eq:op sol pi interaction intermidiate}
      \widehat{\Pi}^{I}(t,x)&=-i\int_{-\infty}^{\infty}\frac{dk}{2\pi}\sqrt{\frac{\omega_{k,\mu}}{2}}
    \left(\widehat{a}^{I}_{k,\mu}(t)e^{ikx}-\widehat{a}^{I \dagger}_{k,\mu}(t)e^{-ikx}\right).
\end{align}
Plugging Eq.~\eqref{eq: a interaction} (and its Hermitian conjugate) into Eqs.~\eqref{eq:op sol phi interaction intermidiate} and \eqref{eq:op sol pi interaction intermidiate} and reorganizing the result as a sum of an operator part and the $c$-number part, we obtain
\begin{align}
    \label{eq:op sol phi interaction}
    \widehat{\phi}^{I}(t,x)
    &= \widehat{\phi}^{I}_{\rm ff}(t,x)+\phi_{\rm cl}(t,x),
    \\
    \label{eq:op sol pi interaction}
     \widehat{\Pi}^{I}(t,x)
     &= \widehat{\Pi}^{I}_{\rm ff}(t,x)+\partial_{t}\phi_{\rm cl}(t,x).
\end{align}
Here we have introduced the free-free parts of the field operators $ \widehat{\phi}^{I}_{\rm ff}(t,x)$ and $ \widehat{\Pi}^{I}_{\rm ff}(t,x)$ which are obtained by evolving $\widehat{\phi}(x)$ and $\widehat{\Pi}(x)$ by the free-free Hamiltonian (\ref{eq:free-free Hamiltonian}):
\begin{align}
    \label{eq:define phi freefree}
    \widehat{\phi}^{I}_{\rm ff}(t,x)
    &= e^{i\widehat{H}_{\rm ff} (t-t_{0})}\widehat{\phi}(x)e^{-i\widehat{H}_{\rm ff} (t-t_{0})}\notag
    \\
    &=\int_{-\infty}^{\infty}\frac{dk}{2\pi\sqrt{2\omega_{k,\mu}}}\left(\widehat{a}_{k,\mu}e^{-i\omega_{k,\mu}(t-t_{0})+ikx}+\widehat{a}^{\dagger}_{k,\mu}e^{i\omega_{k,\mu}(t-t_{0})-ikx}\right),
    \\
     \label{eq:define pi freefree}
    \widehat{\Pi}^{I}_{\rm ff}(t,x)&
    =e^{i\widehat{H}_{\rm ff} (t-t_{0})}\widehat{\Pi}(x)e^{-i\widehat{H}_{\rm ff} (t-t_{0})}\notag
    \\
    &=-i\int_{-\infty}^{\infty}\frac{dk}{2\pi}\sqrt{\frac{\omega_{k,\mu}}{2}}\left(\widehat{a}_{k,\mu}e^{-i\omega_{k,\mu}(t-t_{0})+ikx}-\widehat{a}^{\dagger}_{k,\mu}e^{i\omega_{k,\mu}(t-t_{0})-ikx}\right) .
\end{align}
The $c$-number function $\phi_{\rm cl}(t,x)$ in Eqs.~\eqref{eq:op sol phi interaction} and \eqref{eq:op sol pi interaction} is given by 
\begin{align}
    \phi_{\rm cl}(t,x)&=-\frac{q}{\sqrt{\pi}}\int_{t_{0}}^{t}ds \int dy E_{C}(s,y)\int_{-\infty}^{\infty} \frac{dk}{2\pi \omega_{k,\mu}}\sin\left(\omega_{k,\mu}(t-s)\right)e^{ik(x-y)}\notag
    \\
    &=-\frac{q}{\sqrt{\pi}}\int_{t_{0}}^{t}ds \int dy G_{\rm ret}(t-s,x-y) E_{C}(s,y),
\end{align}
where $G_{\rm ret}(t-s,x-y) $ is the retarded Green function given by (e.g. \cite{Dvornikov:2004xh,Finster.Paganini2023feb}) 
\begin{align}
     G_{\rm ret}(\Delta t,\Delta x)&=\frac{\theta(\Delta t)}{\pi}\int^{\infty}_{-\infty}\!\frac{dk}{2  \omega_{k,\mu}}\sin{(\omega_{k,\mu}\Delta t)}e^{ik\Delta x}\notag
    \\
    &=\frac{1}{2}\theta\left(\Delta t\right)\theta\left(\Delta t^{2}-\Delta x^{2}\right)J_{0}\left(\mu\sqrt{\Delta t^{2}-\Delta x^{2}}\right).
\end{align}
Therefore, $\phi_{\rm  cl}(t,x)$ satisfies the Klein-Gordon equation
\begin{align}
    \label{eq:phi classical}
    \left(\partial_{t}^{2}-\partial_{x}^{2}+\mu^{2}\right)\phi_{\rm cl}(t,x)=-\frac{q}{\sqrt{\pi}}E_{C}(t,x),
\end{align}
with the condition $\phi_{\rm cl}(t,x)=0=\partial_{t}\phi_{\rm cl}(t,x)$ for $t\leq t_{0}$. 

The fact that the field operators are given as a sum of the genuine operator parts $\widehat{\phi}^{I}_{\rm ff}(t,x)$ and $ \widehat{\Pi}^{I}_{\rm ff}(t,x)$, and a $c$-number part $\phi_{\rm cl}(t,x)$ gives rise to an important identity extensively used in the rest of this paper. 
Consider a composite operator $F(\widehat{\phi}^{I}(t,x),\widehat{\Pi}^{I}(t,x))$ defined polynomially in $\widehat{\phi}^{I}(t,x)$ and $\widehat{\Pi}^{I}(t,x)$, and take its normal ordering $N_{\mu}\left\{\cdots\right\}$ with respect to the  Schr{\"o}dinger picture creation and annihilation operators $\widehat{a}_{k,\mu}^{\dagger}$ and $\widehat{a}_{k,\mu}$. The expectation value of this operator in the state $\ket{0,\mu}$ is then given by
\begin{align}
    \label{eq:op to cl relatio}
    &\bra{0,\mu}N_{\mu}\left\{F\left(\widehat{\phi}^{I}(t,x),\widehat{\Pi}^{I}(t,x)\right)\right\}\ket{0,\mu}=
    F\left(\phi_{\rm cl}(t,x),\partial_{t}\phi_{\rm cl}(t,x)\right).
\end{align}
This equality holds since the normal ordering ensures that any term containing $\widehat{a}_{k,\mu}$ and $\widehat{a}^{\dagger}_{k,\mu}$ vanishes when sandwiched with $\ket{0,\mu}$, and the only surviving part is the $c$-number part of $\widehat{\phi}^{I}(t,x)$ and $\widehat{\Pi}^{I}(t,x)$, which are $\phi_{\rm cl}(t,x)$ and $\partial_{t}\phi_{\rm cl}(t,x)$.

\subsubsection{Current}
We now derive an expression for the current expectation value in the interaction picture. 
We have already found the interaction field operators in Eqs.~\eqref{eq:op sol phi interaction} and \eqref{eq:op sol pi interaction}, but we also need the interaction Hamiltonian $\widehat{H}^{I}_{\rm int}(t)$ in the interaction picture,
\begin{align}
    \label{eq: from SchHint to intHint}
    \widehat{H}^{I}_{\rm int}(t)
    &= \int \!dx\,\left(-\frac{e^{\gamma}mf(t)\mu}{2\pi}\right)\widehat{U}_{0}^{\dagger}(t,t_{0})
    N_{\mu}\left\{
   \cos{\sqrt{4\pi}\widehat{\phi}(x)}
    \right\}
    \widehat{U}_{0}(t,t_0).
\end{align}
To fully move to the interaction picture, Eq.~\eqref{eq: from SchHint to intHint} needs to be written solely in terms of the interaction field operators $\widehat{\phi}^{I}(t,x)$. 
To further compute the RHS of Eq.~\eqref{eq: from SchHint to intHint}, we use a special property of normal ordering that arises from Eq.~\eqref{eq: a interaction}. From Eq.~\eqref{eq: a interaction}, one can see that the creation and annihilation operators \textit{do not mix}, i.e., $\widehat{a}_{k,\mu}^{I}(t)$ is given by $\widehat{a}_{k,\mu}$ multiplied by a phase factor $e^{-i\omega_{k,\mu}(t-t_{0})}$ plus a $c$-number function, and it does not depend on $\widehat{a}^{\dagger}_{k,\mu}$. This in fact leads to the preservation of normal ordering. For instance, consider an operator $N_{\mu}\left\{\widehat{a}_{k,\mu}\widehat{a}_{k,\mu}^{\dagger}\right\}$:
\begin{align}
    \label{eq: normal ordering aadagger}
     \widehat{U}_{0}^{\dagger}(t,t_{0})N_{\mu}\left\{\widehat{a}_{k,\mu}\widehat{a}_{k,\mu}^{\dagger}\right\} \widehat{U}_{0}(t,t_{0})
     &=\widehat{U}_{0}^{\dagger}(t,t_{0})\widehat{a}_{k,\mu}^{\dagger}\widehat{a}_{k,\mu}\widehat{U}_{0}(t,t_{0})\notag
     \\
     &=\widehat{U}_{0}^{\dagger}(t,t_{0})\widehat{a}_{k,\mu}^{\dagger}\widehat{U}_{0}(t,t_{0})\widehat{U}_{0}^{\dagger}(t,t_{0})\widehat{a}_{k,\mu}\widehat{U}_{0}(t,t_{0})\notag
     \\
     &=a^{I \dagger}_{k,\mu}(t)a^{I }_{k,\mu}(t)\notag
     \\
     &=N_{\mu}\left\{a^{I }_{k,\mu}(t)a^{I \dagger}_{k,\mu}(t)\right\},
\end{align}
where $a^{I }_{k,\mu}(t)$ is given by Eq.~\eqref{eq: a interaction} with $a^{I \dagger}_{k,\mu}(t)$ its  Hermitian conjugate.  Importantly, the normal ordering in the final line of Eq.~\eqref{eq: normal ordering aadagger} is for the  Schr{\"o}dinger picture operators  $a_{k,\mu}$ and $a^{\dagger}_{k,\mu}$. 
The same argument leading to Eq.~\eqref{eq: normal ordering aadagger} can be used for more general operators.
In particular, for any composite operators $F(\widehat{\phi}(x),\widehat{\Pi}(x))$ defined polynomially in the  Schr{\"o}dinger picture operators  $\widehat{\phi}(x)$ and $\widehat{\Pi}(x)$, we have
\begin{align}\label{notobvious}
    \widehat{U}_{0}^{\dagger}(t,t_{0})N_{\mu}\left\{F(\widehat{\phi}(x),\widehat{\Pi}(x))\right\}\widehat{U}_{0}(t,t_{0})
    =N_{\mu}\left\{F(\widehat{\phi}^{I}(t,x),\widehat{\Pi}^{I}(t, x))\right\}.
\end{align}
In other words, normal ordering commutes with moving to the interaction picture.  
Again, it is worth emphasizing that Eq.~\eqref{notobvious} is a consequence of Eq.~\eqref{eq: a interaction}, i.e., the time evolution by $\widehat{U}_{0}(t,t_0)$ of $\widehat{a}_{k,\mu}$ and $\widehat{a}_{k,\mu}^{\dagger}$ does not mix them.

Applying Eq.~\eqref{notobvious} to Eq.~\eqref{eq: from SchHint to intHint}, the interaction Hamiltonian in the interaction picture is readily given by
\begin{align}
    \label{eq:int hamiltonian int pic}
    \widehat{H}^{I}_{\rm int}(t)=\int dx\, \left(-\frac{e^{\gamma}mf(t)}{2\pi}\right)\mu N_{\mu}\left\{\cos{\sqrt{4\pi}\widehat{\phi}^{I}(t,x)}\right\}.
\end{align}
Now, we are ready to write down the expression of the current expectation value in the interaction picture. 
First, notice that the time evolution of the (Schr\"odinger-picture) state is given by
\begin{align}
\ket{\psi(t)}&\simeq\widehat{U}_{\mu}(t,t_{0})\ket{\Omega}\simeq\widehat{U}_{0}(t,t_{0})\widehat{U}_{\rm int}(t,-\infty)\ket{0,\mu}.
\end{align}
The first equality is the $\mu$-deformed version of \eqref{eq:verysimple}, and the second folows from Eqs.~\eqref{eq:sG vacuum int} and \eqref{eq:define Uint(t,t')} (with $\widehat{U}_T=\widehat{U}_\mu)$.  Restoring the $\mu \to 0$ limit, the current $\braket{j^{x}}$ \eqref{eq:<j> definition} may now be written
\begin{align}
    \label{eq:<j> int picture2}
    \braket{j^{x}}&=-\frac{q}{\sqrt{\pi}}\lim_{\mu\to0}\bra{0,\mu}\widehat{U}_{\rm int}^{\dagger}(t,-\infty)\widehat{\Pi}^{I}(t,x)\widehat{U}_{\rm int}(t,-\infty)\ket{0,\mu},
\end{align}
where $\widehat{U}_{\rm int}(t,t')$ is given by Eq.~(\ref{eq:Uint dyson series}) with the corresponding interaction Hamiltonian given by Eq.~\eqref{eq:int hamiltonian int pic}.

\subsection{Perturbation expansion}
Using Eqs.~(\ref{eq:<j> int picture2}), \eqref{eq:Uint dyson series} and the formula from the ``in-in formalism" \cite{Weinberg:2005vy}, $\braket{j^{x}}$ can be expanded as
\begin{align}
    \label{eq:<j> boson expansion}
    \braket{j^{x}}=&-\frac{q}{\sqrt{\pi}} \sum_{N=0}^{\infty}i^{N}\int_{-\infty}^{t}dt_{N}\int^{t_{N}}_{-\infty}dt_{N-1}\cdots\int^{t_{2}}_{-\infty}dt_{1}\notag
    \\
    &\times\lim_{\mu\to0}\bra{0,\mu}\left[\widehat{H}_{\rm int}^{I}(t_{1}),\left[\widehat{H}_{\rm int}^{I}(t_{2}),\cdots\left[\widehat{H}_{\rm int}^{I}(t_{N}),\widehat{\Pi}^{I}(t,x)\right]\cdots\right]\right]\ket{0,\mu},
\end{align}
where the $N=0$ term should be understood as $\bra{0,\mu} \widehat{\Pi}^{I}(t,x)\ket{0,\mu}$. 
To proceed further, we rewrite Eq.~(\ref{eq:<j> boson expansion}) in a suitable way for later calculations. 
By explicitly writing the expression of the interaction Hamiltonian in the innermost commutator of Eq.~(\ref{eq:<j> boson expansion}) using Eq.~(\ref{eq:int Hamiltonian}), and using Eq.~(\ref{eq:cos pi commu interaction}) to compute the commutator, we have
\begin{align}
     \braket{j^{x}}&=-\frac{q}{\sqrt{\pi}}\sum_{N=0}^{\infty}i^{N}\left(-\frac{ e^{\gamma} m}{2\pi}\right)\int_{-\infty}^{t}f(t_{N})dt_{N}\int dy_{N}\int^{t_{N}}_{-\infty}dt_{N-1}\int dy_{N-1}\cdots\int^{t_{2}}_{-\infty}dt_{1}\int dy^{1}\notag
     \\
     &\times\lim_{\mu\to0}\bra{0,\mu}\left[\widehat{H}^{I}_{\rm int}(t_{1}),\left[\widehat{H}^{I}_{\rm int}(t_{2}),\cdots\left[\mu N_{\mu}\left\{\cos{\sqrt{4\pi}\widehat{\phi}^{I}(t_{N},y_{N})}\right\},\widehat{\Pi}^{I}(t,x)\right]\cdots\right]\right]\ket{0,\mu}
    \\
    =&-\frac{q}{\sqrt{\pi}} \sum_{N=0}^{\infty}i^{N}\int_{-\infty}^{t}dt_{N}\int dy_{N}
    (-\sqrt{4\pi})\notag
    \\
    &\times\left[\widehat{\phi}^{I}(t_{N},y_{N}),\widehat{\Pi}^{I}(t,x)\right]\left(-\frac{ e^{\gamma} mf(t_{N})}{2\pi}\right)
   \int^{t_{N}}_{-\infty}dt_{N-1}\cdots\int^{t_{2}}_{-\infty}dt_{1}\notag
     \\
     &\times\lim_{\mu\to0}\bra{0,\mu}\left[\widehat{H}^{I}_{\rm int}(t_{1}),\left[\widehat{H}^{I}_{\rm int}(t_{2}),\cdots\left[\widehat{H}^{I}_{\rm int}(t_{N-1}),\mu N_{\mu}\left\{\sin{\sqrt{4\pi}\widehat{\phi}^{I}(t_{N},y_{N})}\right\}\right]\cdots\right]\right]\ket{0,\mu}
\end{align}
By separating the $N=0$ case and renaming $t_{N}=T$ and $y_{N}=Y$, we can collapse the infinite sum in the following way:
\begin{align}
    \label{eq:j expectation value boson}
    \braket{j^{x}}&= 
     -\frac{q}{\sqrt{\pi}} \partial_{t}\phi_{\rm cl}(t,x)\notag
     \\
     &\quad-q \int_{-\infty}^{t} dT \int_{-\infty}^{\infty} dY
      i\left[\widehat{\phi}^{I}(T,Y),\widehat{\Pi}^{I}(t,x)\right]\left(\frac{ e^{\gamma} mf(T)}{\pi}\right)\Braket{\mu N_{\mu}\left\{\sin{\sqrt{4\pi}\phi(T,Y)}\right\}},
\end{align}
where for the first term of the RHS, the $\mu\to0$ limit is understood to have already been taken (i.e., $\phi_{\rm cl}$ solves Eq.~(\ref{eq:phi classical}) with $\mu=0$).
For the second term of Eq.~(\ref{eq:j expectation value boson}), we introduced the notation 
\begin{align}
    \nonumber
    &\Braket{\mu N_{\mu}\left\{\sin{\sqrt{4\pi}\phi(T,Y)}\right\}}
    \\
     \label{eq<sin>boson2} 
    &=\lim_{\mu\to0}\bra{0,\mu}\widehat{U}^{\dagger}_{\rm int}(T,-\infty)\mu N_{\mu}\left\{\sin{\sqrt{4\pi}\widehat{\phi}^{I}(T,Y)}\right\}\widehat{U}_{\rm int}(T,-\infty)\ket{0,\mu}.
\end{align}
In order to determine the leading mass correction to the current, we must evaluate the second term in Eq.~(\ref{eq:j expectation value boson}) to leading order in  $m$.  At $O(m^0)$ we have $\widehat{U}_{\rm int}=\widehat{1}$, so Eq.~\eqref{eq<sin>boson2} becomes
\begin{align}
    \Braket{\mu N_{\mu}\left\{\sin{\sqrt{4\pi}\phi(T,Y)}\right\}}\Big|_{m=0} & = \lim_{\mu\to0}\bra{0,\mu}\mu N_{\mu}\left\{\sin{\sqrt{4\pi}\widehat{\phi}^{I}(T,Y)}\right\}\ket{0,\mu}\Big|_{m=0} \\
    & = \lim_{\mu\to0}\mu \sin{\sqrt{4\pi}\phi_{\rm cl}(T,Y)}\Big|_{m=0}= 0,
\end{align}
with the second-to-last step following from Eq.~\eqref{eq:op to cl relatio}.  We must therefore proceed to the next order in $m$.  We will first expand naively in $m$ but encounter an infinity.  We will then circle back and discover a missing $m \log m$ contribution; making the corrected ansatz provides finite results.

\subsubsection{Failure of naive expansion}\label{sec:failure naive expansion}
We now illustrate what goes wrong if one simply expands Eq.~\eqref{eq<sin>boson2} in $m$ in the usual way.  Using Eqs.~(\ref{eq:op to cl relatio}) and (\ref{eq:cos sin commu}), we find
\begin{align}
    \label{eq:<sin> expand boson naive}
    & \Braket{\mu N_{\mu}\left\{\sin{\sqrt{4\pi}\phi(T,Y)}\right\}}\notag
    \\
    &=\int_{-\infty}^{T}dt_{1}\int dy_{1}
    i\left(-\frac{e^{\gamma}mf(t_1)}{2\pi}\right) \notag
    \\
    &\hspace{2cm}\times\lim_{\mu\to0}\bra{0,\mu}
    \left[
    \mu N_{\mu}\left\{\cos{\sqrt{4\pi}\widehat{\phi}^{I}(t_{1},y_{1})}\right\}
    ,\mu N_{\mu}\left\{\sin{\sqrt{4\pi}\widehat{\phi}^{I}(T,Y)}\right\}\right]
    \ket{0,\mu}\notag
    \\
    &=
     -\frac{2m}{e^{\gamma}} \int_{-\infty}^{T}dt_{1}f(t_1)\int dy_{1}
    \sin{\left[\sqrt{4\pi}\phi_{\rm cl}(t_{1},y_{1})-\sqrt{4\pi}\phi_{\rm cl}(T,Y)\right]}\delta\left(\Delta t^{2} -\Delta y^{2}\right),\notag
    \\
    &= \frac{2m}{e^{\gamma}} \int_{-\infty}^{T}\frac{f(t_1)dt_{1}}{2(T-t_{1})}\left\{
     \sin{\left[\sqrt{4\pi}\phi_{\rm cl}(T,Y)-\sqrt{4\pi}\phi_{\rm cl}(t_{1},Y-T+t_{1})\right]}\right.\notag
     \\
     &\hspace{3cm}+\left.
      \sin{\left[\sqrt{4\pi}\phi_{\rm cl}(T,Y)-\sqrt{4\pi}\phi_{\rm cl}(t_{1},Y+T-t_{1})\right]}
    \right\}
\end{align}
where $\Delta t=T-t_{1}$ and $\Delta y=Y-y_{1}$. 
When $T<t_0$, we have $t_1<T<t_0$ and both $\phi_{\rm cl}(T,Y)$ and $\phi_{\rm cl}(t_1,Y\mp T\pm t_1)$ are evaluated before the electric field turns on; thus Eq.~\eqref{eq:<sin> expand boson naive} vanishes entirely.
For $T>t_0$, however, this integral is divergent. To show this, first write the integral in Eq.~\eqref{eq:<sin> expand boson naive} as a sum of the integral from $-\infty$ to $t_{0}$ and the integral from $t_{0}$ to $T$:
\begin{align}
    \int_{-\infty}^{T}dt_{1}=\int_{-\infty}^{t_{0}}dt_{1}+\int_{t_{0}}^{T}dt_{1}.
\end{align}
For the $(-\infty,t_{0}]$ integral,  $\phi_{\rm cl}(t_{1},Y-T+t_{1})$ and $\phi_{\rm cl}(t_{1},Y+T-t_{1})$ vanish since $t_1<t_0$. In addition, since $\phi_{\rm cl}(T,Y)$ does not depend on $t_{1}$, 
$\sin{\left[\sqrt{4\pi}\phi_{\rm cl}(T,Y)\right]}$ can be pulled out of the integral. 
Then, the $(-\infty,t_{0}]$ part of the integral~\eqref{eq:<sin> expand boson naive} becomes
\begin{align}
    \label{eq:diverging integral}
    \frac{2m}{e^{\gamma}}\sin{\left[\sqrt{4\pi}\phi_{\rm cl}(T,Y)\right]}\int_{-\infty}^{t_{0}}\frac{f(t_1)dt_{1}}{(T-t_{1})}.
\end{align}
Recall that the function $f(t)$ is introduced to adiabatically define the vacuum and has the property that its time derivative approaches $0$ in the adiabatic limit. Thus, in this limit, not only does $f(t)$ approach $1$, but also the region over which $f(t)\sim 1$ extends increasingly far into the past. Consequently,
the integrand in Eq.~\eqref{eq:diverging integral} approaches $1/(T-t_1)$ over an increasingly large time interval, producing a logarithmic divergence associated with the infinite past, $t_1\to-\infty$ \footnote{
To see this more explicitly, consider $f(t)=F[\varepsilon(t_0-t)]$ with $F(0)=1$ and $F(x)\to 0$ as $x\to\infty$. Then, the time scale $1/\varepsilon$ corresponds to the switching time scale, and $\varepsilon\to0$ is the adiabatic limit. With the change of variable $x=\varepsilon(t_0-t_1)$ and defining $\Delta=T-t_{0}$, the integral part of Eq.~\eqref{eq:diverging integral} becomes 
\begin{align*}
    \int_{-\infty}^{t_0}\frac{f(t_1)}{T-t_1}\,dt_1
    &=\int_0^\infty\frac{F(x)}{x+\varepsilon \Delta}\,dx
    \\
    &=
    \int_0^1 \frac{dx}{x+\varepsilon\Delta}
    +\int_0^1 \frac{F(x)-1}{x+\varepsilon\Delta}\,dx
    +\int_1^\infty \frac{F(x)}{x+\varepsilon\Delta}\,dx
    \\
    &=
    -\ln\!\left(\varepsilon \Delta\right)+O(1).
\end{align*}
Thus, in the adiabatic limit $\varepsilon\to0$, Eq.~\eqref{eq:diverging integral} diverges logarithmically with the switching time scale $1/\varepsilon$.  
}. 
Note that since we assumed $T>t_0$, the integral range of Eq.~$\eqref{eq:diverging integral}$ does not contain the diverging part of $1/(T-t_1)$; thus the divergence of Eq.~\eqref{eq:diverging integral} is unrelated to the divergence of $1/(T-t_1)$ at $t_1=T$. 

The $[t_0,T]$ part of the integral~\eqref{eq:<sin> expand boson naive}, on the other hand, is finite even though the integral contains a factor $1/(T-t_1)$ which apparently diverges at $t_1=T$. 
To see this, let $\delta=T-t_1$ and consider the region of small $\delta$ (i.e., $t_1\sim T$). Then, assuming that the spatial and temporal derivatives of $\phi_{\rm cl}$ are finite at $(T,Y)$, the argument of sine in Eq.~\eqref{eq:<sin> expand boson naive} becomes 
\begin{align}
    \sqrt{4\pi}\phi_{\rm cl}(T,Y)-\sqrt{4\pi}\phi_{\rm cl}(T-\delta,Y-\delta)&= \sqrt{4\pi}(\partial_T+\partial_Y)\phi_{\rm cl}(T,Y) \delta+O(\delta^2),
    \\
    \sqrt{4\pi}\phi_{\rm cl}(T,Y)-\sqrt{4\pi}\phi_{\rm cl}(T-\delta,Y+\delta)&= \sqrt{4\pi}(\partial_T-\partial_Y)\phi_{\rm cl}(T,Y) \delta+O(\delta^2).
\end{align}
Therefore, each sine term in Eq.~\eqref{eq:<sin> expand boson naive} behaves $O(T-t_1)$ as $t_1\to T$. This cancels the $1/(T-t_1)$ factor and makes the  $[t_0,T]$ part finite. 
Since one part is infinite and the other part is finite, the total integral \eqref{eq:<sin> expand boson naive} is divergent.

\subsubsection{Corrected perturbation expansion}
The divergence of Eq.~\eqref{eq:<sin> expand boson naive} signals that a smooth Taylor expansion in $m$ is not applicable.  In fact, we will see that the expansion of Eq.~\eqref{eq<sin>boson2} starts at $m\log{m}$ rather than $m$.   
To sort this out, we first note that the divergence of Eq.~\eqref{eq:<sin> expand boson naive} comes from the fact that the range of the time integral is from $t_{1}\to -\infty$ (infinite past), whose appearance is traced back to $\widehat{U}_{\rm int}(T,-\infty)$ in Eq.~\eqref{eq<sin>boson2}. 
To isolate the contribution from the infinite past, we write $\widehat{U}_{\rm int}(T,-\infty)$ as
\begin{align}
    \widehat{U}_{\rm int}(T,-\infty)=\widehat{U}_{\rm int}(T,t_{0})\widehat{U}_{\rm int}(t_{0},-\infty).
\end{align}
Then, using Eq.~\eqref{eq:sG vacuum int}, Eq.~\eqref{eq<sin>boson2} can be rewritten as
\begin{align}
     \Braket{\mu N_{\mu}\left\{\sin{\sqrt{4\pi}\phi(T,Y)}\right\}}
     \label{eq:<sin>boson}
     &\equiv\lim_{\mu\to0}\bra{\Omega}\widehat{U}^{\dagger}_{\rm int}(T,t_{0})\mu N_{\mu}\left\{\sin{\sqrt{4\pi}\widehat{\phi}^{I}(T,Y)}\right\}\widehat{U}_{\rm int}(T,t_{0})\ket{\Omega}.
\end{align}
The key idea is that instead of expanding $\widehat{U}_{\rm int}(t,-\infty)$ in Eq.~\eqref{eq<sin>boson2}, we expand $\widehat{U}_{\rm int}(T,t_{0})$ in Eq.~(\ref{eq:<sin>boson}) in $m$, while keeping the true vacuum $\ket{\Omega}$ untouched.  Thus cuts off the integral at time $t_{0}$, avoiding the divergence inherited from the infinite past. 
Using Eq.~\eqref{eq:define Uint(t,t')}
to expand $\widehat{U}_{\rm int}(T,t_{0})$  to first order in $m$, Eq.~(\ref{eq:<sin>boson}) becomes 
\begin{align}
    \label{eq:<sin> expand boson}
    \Braket{\mu N_{\mu}\left\{\sin{\sqrt{4\pi}\phi(T,Y)}\right\}}=  P_1+P_2,
\end{align}
where 
\begin{align}\label{eq:P1}
    P_1=
    \lim_{\mu\to0}\bra{\Omega}
    \mu N_{\mu}\left\{\sin{\sqrt{4\pi}\widehat{\phi}^{I}(T,Y)}\right\}
    \ket{\Omega}
\end{align}
and
\begin{align}
    \label{eq: define 2nd term}
    &P_2= \lim_{\mu\to0}\int_{t_{0}}^{T}dt_{1}\int dy_{1}
    i\left(-\frac{e^{\gamma}m}{2\pi}\right)\notag \\
    &\qquad\times\bra{\Omega}
    \left[
    \mu N_{\mu}\left\{\cos{\sqrt{4\pi}\widehat{\phi}^{I}(t_{1},y_{1})}\right\}
    ,\mu N_{\mu}\left\{\sin{\sqrt{4\pi}\widehat{\phi}^{I}(T,Y)}\right\}\right]
    \ket{\Omega}.
\end{align}

In Eq.~\eqref{eq: define 2nd term}, we set $f(t_{1})=1$ since $t_{0}<t_{1}<T$ over the range of the integral, so the interaction is already turned on. One can see that Eq.~\eqref{eq: define 2nd term} is very similar to Eq.~\eqref{eq:<sin> expand boson naive} except that the $t_1$ integral now begins at $t_0$ (instead of $-\infty$), no $f(t_1)$ is present, and the state is $\ket{\Omega}$ rather than $\ket{0,\mu}$.  However, the latter two states actually agree at leading order (see \eqref{eq:sG vacuum int} noting that $\widehat{U}_{\rm int}(t,t')=1$ to zeroth order in $m$), so we may replace $\ket{\Omega}$ with $\ket{0,\mu}$ consistent to order $m$ in \eqref{eq: define 2nd term}. Then, using a similar calculation leading to Eq.~\eqref{eq:<sin> expand boson naive}, we obtain
\begin{align}
    \label{eq:<sin> perturbative piece}
    &P_2
    = \frac{2m}{e^{\gamma}} \int_{t_{0}}^{T}\frac{dt_{1}}{2(T-t_{1})}\left\{
     \sin{\left[\sqrt{4\pi}\phi_{\rm cl}(T,Y)-\sqrt{4\pi}\phi_{\rm cl}(t_{1},Y-T+t_{1})\right]}\right.\notag
     \\
     &\hspace{4cm}+\left.
      \sin{\left[\sqrt{4\pi}\phi_{\rm cl}(T,Y)-\sqrt{4\pi}\phi_{\rm cl}(t_{1},Y+T-t_{1})\right]}
    \right\}.
\end{align}
We emphasize that Eq.~\eqref{eq:<sin> perturbative piece} is finite: since the integral range is from $t_0$, the divergence associated with the infinite past has been removed. 

Since $P_2$ is manifestly finite at this order, the divergence we found must originate from $P_1$.  That is, it would not be appropriate to simply expand $P_1$ in $m$. However, from Eq.~(\ref{eq:op sol phi interaction}) and the sine addition formula, we may express $P_1$ \eqref{eq:P1} as
\begin{align}
    \label{eq:sin a+b=sin a con b+ cos a sin b}
    &P_1=
    \sin{\sqrt{4\pi}\phi_{\rm cl}(T,Y)}\lim_{\mu\to0}\bra{\Omega}
    \mu N_{\mu}\left\{\cos{\sqrt{4\pi}\widehat{\phi}^{I}_{\rm ff}(T,Y)}\right\}
    \ket{\Omega}\notag
    \\
    &\hspace{1cm}+
    \cos{\sqrt{4\pi}\phi_{\rm cl}(T,Y)}\lim_{\mu\to0}\bra{\Omega}
    \mu N_{\mu}\left\{\sin{\sqrt{4\pi}\widehat{\phi}^{I}_{\rm ff}(T,Y)}\right\}
    \ket{\Omega}.
\end{align}
This expression is promising because each term on the RHS factors into a c-number function of the classical field $\phi_{\rm cl}$ and a quantum expectation value that is \textit{independent of the background field}.  These expectation values are difficult to compute on the boson side of the duality since they involve the true vacuum $\ket{\Omega}$ rather than the Fock vacuum $\ket{0,\mu}$.  However, on the fermion side the true vacuum \textit{does} coincide with the (massive) Fock vacuum $\ket{0_{\rm f},m}$ \eqref{eq: fermi vacuum = boson vacuum}, allowing us to calculate these expectation values with free fermion theory.  This calculation will reveal that $P_1$ scales as $m\log m$, explaining the divergence that appeared when it was naively expanded in $m$.

\subsubsection{Calculation of $P_1$ using free fermion theory}
We now map the expectation values in \eqref{eq:sin a+b=sin a con b+ cos a sin b} to quantities that can be calculated in free fermion theory without a background field and thereby determine $P_1$.  The relevant Hamiltonian is (\ref{eq: fermion quantum hamiltonian}) without the background field, which we denote by $\widehat{H}_{E_{C}=0}$.  Our goal is to calculate perturbatively in $m$, which requires us to treat the mass term as an interaction.  We therefore write $\widehat{H}_{E_{C}=0}$ as
\begin{align}
    \label{eq:fermi H=Hff+Hint}
    \widehat{H}_{E_{C}=0}=\widehat{H}_{\rm ff}+\widehat{H}_{\rm int},
\end{align}
where
\begin{align}
    \label{eq:define fermi freefree Hamiltonian}
    \widehat{H}_{\rm ff}=&\int dx :\overline{\widehat{\psi}}(x)\left(-i\gamma^{x}\partial_{x}\right)\widehat{\psi}(x):_{0},
    \\
    \label{eq:define fermi int Hamiltonian}
    \widehat{H}_{\rm int}=&m\int dx  :\overline{\widehat{\psi}}(x)\widehat{\psi}(x):_{0}.
\end{align}
We chose the normal-ordering mass for fermions to be zero, which is denoted by the symbol $:\cdots:_{0}$. This choice is convenient since the operator $\widehat{\phi}^{I}_{\rm ff}$ in Eq.~\eqref{eq:sin a+b=sin a con b+ cos a sin b} is evolved by the free-free Hamiltonian \eqref{eq:free-free Hamiltonian} which corresponds to the massless fermion Hamiltonian. 

We denote the massless annihilation (and creation) operators by $\widehat{B}_{k}$ and $\widehat{D}_{k}$ and define them by expanding the fermion field $\widehat{\psi}(x)$ as
\begin{align}
    \label{eq:psi define B and D}
    \widehat{\psi}(x)=\begin{pmatrix}
        \widehat{\psi}_{1}(x)\\
        \widehat{\psi}_{2}(x)
    \end{pmatrix}
    =&\int_{-\infty}^{\infty}\frac{dk}{2\pi}
    \left[
    \widehat{B}_{k}e^{ikx}\begin{pmatrix}
        \theta(-k)\\
        \theta(k)
    \end{pmatrix}
    +\widehat{D}^{\dagger}_{k}e^{-ikx}\begin{pmatrix}
        -\theta(-k)\\
       \theta(k)
    \end{pmatrix}
    \right].
\end{align}
Then, the massless normal-ordering $:\cdots:_{0}$ places all $\widehat{B}_{k}$ and $\widehat{D}_{k}$ to the right of all $\widehat{B}^{\dagger}_{k}$ and $\widehat{D}^{\dagger}_{k}$. 
Similar to defining the free-free interaction picture operators (\ref{eq:define phi freefree}) and (\ref{eq:define pi freefree}) on the boson side, we can define the free-free interaction picture operator $\widehat{\psi}_{\rm ff}^{I}(t,x)$ on the fermion side by
\begin{align}
    \widehat{\psi}_{\rm ff}^{I}(t,x)=e^{i\widehat{H}_{\rm ff}(t-t_{0})}\widehat{\psi}(x)e^{-i\widehat{H}_{\rm ff}(t-t_{0})}.
\end{align}
On account of the choice of the massless normal-ordering, the free-free interaction operator is given in terms of the massless creation and annihilation operators by
\begin{align}
    \label{eq:psi define B and D interaction}
    \widehat{\psi}^{I}_{\rm ff}(t,x)&=\begin{pmatrix}
        \widehat{\psi}_{1\rm ff}^{I}(t,x)\\
        \widehat{\psi}_{2\rm ff}^{I}(t,x)
    \end{pmatrix}\notag
    \\
    &=\int_{-\infty}^{\infty}\frac{dk}{2\pi}
    \left[
    \widehat{B}_{k}e^{-i|k|(t-t_{0})+ikx}\begin{pmatrix}
        \theta(-k)\\
        \theta(k)
    \end{pmatrix}
    +\widehat{D}^{\dagger}_{k}e^{i|k|(t-t_{0})-ikx}\begin{pmatrix}
        -\theta(-k)\\
       \theta(k)
    \end{pmatrix}
    \right].
\end{align}
In addition, since evolution by the $e^{-i\widehat{H}_{\rm ff}(t-t_0)}$ is a free evolution, we have
\begin{align}
     e^{i\widehat{H}_{\rm ff}(t-t_{0})}\widehat{B}_{k}e^{-i\widehat{H}_{\rm ff}(t-t_{0})}&=\widehat{B}_{k}e^{-i|k|(t-t_{0})},
     \\
     e^{i\widehat{H}_{\rm ff}(t-t_{0})}\widehat{D}_{k}e^{-i\widehat{H}_{\rm ff}(t-t_{0})}&=\widehat{D}_{k}e^{-i|k|(t-t_{0})}.
\end{align}
Thus, no mixing of the creation and annihilation operators by this evolution, leading to the commutation of the massless normal-ordering $:\cdots:_{0}$ and time evolution by $e^{-i\widehat{H}_{\rm ff}(t-t_0)}$. Therefore, we have 
\begin{align}
    \label{eq: bar psi psi free free interaction}
    e^{i\widehat{H}_{\rm ff}(t-t_{0})}:\overline{\widehat{\psi}}(x)\widehat{\psi}(x):_{0}e^{-i\widehat{H}_{\rm ff}(t-t_{0})}&=:\overline{\widehat{\psi}^{I}}_{\rm ff}(t,x)\widehat{\psi}^{I}_{\rm ff}(t,x):_{0},
    \\
    \label{eq: bar psi gamma5 psi free free interaction}
    e^{i\widehat{H}_{\rm ff}(t-t_{0})}:\overline{\widehat{\psi}}(x)\gamma^{5}\widehat{\psi}(x):_{0}e^{-i\widehat{H}_{\rm ff}(t-t_{0})}&=:\overline{\widehat{\psi}^{I}}_{\rm ff}(t,x)\gamma^{5}\widehat{\psi}^{I}_{\rm ff}(t,x):_{0}.
\end{align}
Now applying the bosonization identities~(\ref{eq:fermi=boson mass}) and (\ref{eq:fermi=boson chiral}) to the LHS of Eqs.~\eqref{eq: bar psi psi free free interaction} and \eqref{eq: bar psi gamma5 psi free free interaction} gives
\begin{align}
    \label{eq: boson fermion mass free free interaction}
    e^{i\widehat{H}_{\rm ff}(t-t_{0})}:\overline{\widehat{\psi}}(x)\widehat{\psi}(x):_{0}e^{-i\widehat{H}_{\rm ff}(t-t_{0})}
    &=  -\frac{e^{\gamma}\mu}{2\pi}e^{i\widehat{H}_{\rm ff}(t-t_{0})}
   N_{\mu}\left\{\cos{\sqrt{4\pi}\widehat{\phi}(x)}\right\}
    e^{-i\widehat{H}_{\rm ff}(t-t_{0})}
    \\
    \label{eq: boson fermion chiral free free interaction}
    e^{i\widehat{H}_{\rm ff}(t-t_{0})}: i\overline{\widehat{\psi}}(x)\gamma^{5}\widehat{\psi}(x):_{0}e^{-i\widehat{H}_{\rm ff}(t-t_{0})}
    &=
       \frac{e^{\gamma}\mu}{2\pi}e^{i\widehat{H}_{\rm ff}(t-t_{0})}
   N_{\mu}\left\{\sin{\sqrt{4\pi}\widehat{\phi}(x)}\right\}
    e^{-i\widehat{H}_{\rm ff}(t-t_{0})}
\end{align}
For the RHS of these expressions, the normal ordering can pass $ e^{i\widehat{H}_{\rm ff}(t-t_{0})}$ since $ e^{i\widehat{H}_{\rm ff}(t-t_{0})}$ does not mix the boson creation and annihilation operators. Therefore, combining this fact with Eqs.~\eqref{eq: bar psi psi free free interaction} and \eqref{eq: bar psi gamma5 psi free free interaction}, we obtain
\begin{align}
     \label{eq:fermi=boson mass free-free interaction}
   :\overline{\widehat{\psi}^{I}}_{\rm ff}(t,x)\widehat{\psi}^{I}_{\rm ff}(t,x):_{0}&=-\frac{e^{\gamma}\mu}{2\pi}N_{\mu}\left\{\cos{\sqrt{4\pi}\widehat{\phi}^{I}_{\rm ff}(t,x)}\right\},
    \\
     \label{eq:fermi=boson chiral free-free interaction}
    i:\overline{\widehat{\psi}^{I}}_{\rm ff}(t, x)\gamma^{5}\widehat{\psi}^{I}_{\rm ff}(t,x):&=\phantom{a}\frac{e^{\gamma}\mu}{2\pi_{0}}N_{\mu}\left\{\sin{\sqrt{4\pi}\widehat{\phi}^{I}_{\rm ff}(t,x)}\right\}.
\end{align}
Taking the expectation value in the vacuum $\ket{\Omega}=\ket{0_{\rm f},m}$ (\ref{eq: fermi vacuum = boson vacuum}) yields
\begin{align}
    \label{eq:cos exact}
    \lim_{\mu\to0}\left(\frac{e^{\gamma}\mu}{2\pi}\right)\bra{\Omega} N_{\mu}\left\{\cos{\sqrt{4\pi}\widehat{\phi}^{I}_{\rm ff}(t,x)}\right\}
    \ket{\Omega}
    &=-\bra{0_{\rm f},m}:\overline{\widehat{\psi}^{I}}_{\rm ff}(t,x)\widehat{\psi}^{I}_{\rm ff}(t,x):_{0}\ket{0_{\rm f},m},
    \\
    \label{eq:sin exact}
    \lim_{\mu\to0}\left(\frac{e^{\gamma}\mu}{2\pi}\right)\bra{\Omega} N_{\mu}\left\{\sin{\sqrt{4\pi}\widehat{\phi}^{I}_{\rm ff}(t,x)}\right\}
    \ket{\Omega}
    &= i\bra{0_{\rm f},m}:\overline{\widehat{\psi}^{I}}_{\rm ff}(t,x)\gamma^{5}\widehat{\psi}^{I}_{\rm ff}(t,x):_{0}\ket{0_{\rm f},m}.
\end{align}
Since the LHS of Eqs.~\eqref{eq:cos exact} and \eqref{eq:sin exact} appear in the expression of $P_1$ \eqref{eq:sin a+b=sin a con b+ cos a sin b}, the evaluation of $P_1$ now reduces to computing the RHS of  Eqs.~\eqref{eq:cos exact} and \eqref{eq:sin exact}, which can be fully done in terms of the free fermion theory. 
A convenient way to compute them is to use the Bogoliubov transformation between $\widehat{b}_{k}, \widehat{d}_{k}$ and $\widehat{B}_{k}, \widehat{D}_{k}$, which is obtained by equating Eq.~(\ref{eq:psi define b and d}) and Eq.~(\ref{eq:psi define B and D}), resulting in
\begin{align}
    \label{eq:bogoliubov transform bd to BD}
    \begin{pmatrix}
        \widehat{B}_{k}
        \\
        \widehat{D}^{\dagger}_{-k}
    \end{pmatrix}
    =&\frac{1}{\sqrt{1+U_{k}^{2}}}
    \begin{pmatrix}
        1 & U_{k}
        \\
        -U_{k} & 1
    \end{pmatrix}
     \begin{pmatrix}
        \widehat{b}_{k}
        \\
        \widehat{d}^{\dagger}_{-k}
    \end{pmatrix},
\end{align}
with
\begin{align}
    \label{eq:Uk bogoliubov }
    U_{k}=&-\frac{k}{m}\left(1-\sqrt{\left(\frac{m}{k}\right)^{2}+1}\right).
\end{align}
Since the operational rules for the action of $\widehat{b}_{k}, \widehat{d}_{k}, \widehat{b}^{\dagger}_{k}$, and $\widehat{d}^{\dagger}_{k}$ on $\ket{0_{\rm f},m}$ are fully specified, we have the following identities using Eq.~(\ref{eq:bogoliubov transform bd to BD}): 
\begin{align}
    \label{eq:<BdaggerB>}
    \bra{0_{\rm f},m}\widehat{B}_{k_{1}}^{\dagger}\widehat{B}_{k_{2}}\ket{0_{\rm f},m}=&\frac{U_{k_{1}}^{2}}{1+U_{k_{1}}^{2}}2\pi\delta(k_{1}-k_{2}),
    \\
    \label{eq:<DDdagger>}
    \bra{0_{\rm f},m}\widehat{D}_{k_{1}}^{\dagger}\widehat{D}_{k_{2}}\ket{0_{\rm f},m}=&\frac{U_{k_{1}}^{2}}{1+U_{k_{1}}^{2}}2\pi\delta(k_{1}-k_{2}),
    \\
    \label{eq:<DB>}
    \bra{0_{\rm f},m}\widehat{D}_{k_{1}}\widehat{B}_{k_{2}}\ket{0_{\rm f},m}=&\frac{-U_{k_{1}}}{1+U_{k_{1}}^{2}}2\pi\delta(k_{1}+k_{2}),
    \\
    \label{eq:<BdaggerDdagger>}
    \bra{0_{\rm f},m}\widehat{B}_{k_{1}}^{\dagger}\widehat{D}_{k_{2}}^{\dagger}\ket{0_{\rm f},m}=&\frac{U_{k_{1}}}{1+U_{k_{1}}^{2}}2\pi\delta(k_{1}+k_{2}).
\end{align}
Then, using the gamma matrices (\ref{eq:gamma matrix}), we have
\begin{align}
    \label{eq:psipsi=psi2psi1+psi1psi2}
    \overline{\widehat{\psi}^{I}}_{\rm ff}(t,x)\widehat{\psi}^{I}_{\rm ff}(t,x)&=
    \overline{\widehat{\psi}^{I}}_{2\rm ff}(t,x)\widehat{\psi}^{I}_{1\rm ff}(t,x)
    +
    \overline{\widehat{\psi}^{I}}_{1\rm ff}(t,x)\widehat{\psi}^{I}_{2\rm ff}(t,x),
    \\
    \label{eq:psi5psi=-psi2psi1+psi1psi2}
     \overline{\widehat{\psi}^{I}}_{\rm ff}(t,x)\gamma^{5}\widehat{\psi}^{I}_{\rm ff}(t,x)&=
    -\overline{\widehat{\psi}^{I}}_{2\rm ff}(t,x)\widehat{\psi}^{I}_{1\rm ff}(t,x)
    +
    \overline{\widehat{\psi}^{I}}_{1\rm ff}(t,x)\widehat{\psi}^{I}_{2\rm ff}(t,x),
\end{align}
where $\widehat{\psi}^{I}_{1\rm ff}(t,x)$ and $\widehat{\psi}^{I}_{2\rm ff}(t,x)$ denote the components of $\widehat{\psi}^{I}_{\rm ff}(t,x)$ has two components as in Eq.~\eqref{eq:psi define B and D interaction}.  Now the RHS of Eqs.~(\ref{eq:psipsi=psi2psi1+psi1psi2}) and (\ref{eq:psi5psi=-psi2psi1+psi1psi2}) can be evaluated piece by piece using Eqs.~(\ref{eq:<BdaggerB>})--(\ref{eq:<BdaggerDdagger>}), which results in 
\begin{align}
    \label{eq:psi2psi1 interaction}
    \bra{0_{\rm f},m}\widehat{\psi}^{I\dagger}_{2\rm ff}(t,x)\widehat{\psi}^{I}_{1\rm ff}(t,x)\ket{0_{\rm f},m}
    &=
     \int_{-\infty}^{\infty}  \frac{dk_{1}}{2\pi}\theta(k_{1})\frac{-2U_{k_{1}}}{1+U_{k_{1}}^{2}}\cos{\left\{2k_{1}(t-t_{0})\right\}}\notag
    \\
    &=-\int_{0}^{\infty}\frac{dk_1}{2\pi}\frac{m}{\sqrt{m^{2}+k_1^{2}}}\cos{\left\{2k_{1}(t-t_{0})\right\}}\notag
    \\
    &=-\frac{m}{2\pi}K_{0}\left(2m(t-t_{0})\right),
\end{align}
where $K_{0}(x)$ is the modified Bessel function of the second kind of order zero. Similarly, we obtain
\begin{align}
     \label{eq:psi1psi2 interaction}
    \bra{0_{\rm f},m}\widehat{\psi}^{I\dagger}_{1\rm ff}(t,x)\widehat{\psi}^{I}_{2\rm ff}(t,x)\ket{0_{\rm f},m}=&-\frac{m}{2\pi}K_{0}\left(2m(t-t_{0})\right).
\end{align}
Plugging Eqs.~(\ref{eq:psi2psi1 interaction}) and (\ref{eq:psi1psi2 interaction}) into Eqs.~(\ref{eq:psipsi=psi2psi1+psi1psi2}) and (\ref{eq:psi5psi=-psi2psi1+psi1psi2}) leads to 
\begin{align}
    \bra{0_{\rm f},m}\overline{\widehat{\psi}^{I}}_{\rm ff}(t,x)\widehat{\psi}^{I}_{\rm ff}(t,x)\ket{0_{\rm f},m}=&-\frac{m}{\pi}K_{0}\left(2m(t-t_{0})\right),
    \\
    \bra{0_{\rm f},m}\overline{\widehat{\psi}^{I}}_{\rm ff}(t,x)\gamma^{5}\widehat{\psi}^{I}_{\rm ff}(t,x)\ket{0_{\rm f},m}=&0.
\end{align}
Thus, using Eqs.~(\ref{eq:cos exact}) and (\ref{eq:sin exact}), we obtain the following exact result: 
\begin{align}
    \label{eq:cos exact2}
    &\lim_{\mu\to0}\left(\frac{e^{\gamma}\mu}{2\pi}\right)\bra{\Omega} N_{\mu}\left\{\cos{\sqrt{4\pi}\widehat{\phi}^{I}_{\rm ff}(t,y)}\right\}
    \ket{\Omega}=\frac{m}{\pi}K_{0}\left(2m(t-t_{0})\right),
    \\
    \label{eq:sin exact2}
    &\lim_{\mu\to0}\left(\frac{e^{\gamma}\mu}{2\pi}\right)\bra{\Omega} N_{\mu}\left\{\sin{\sqrt{4\pi}\widehat{\phi}^{I}_{\rm ff}(t,y)}\right\}
    \ket{\Omega}=0.
\end{align}
Plugging Eqs.~(\ref{eq:cos exact2}) and (\ref{eq:sin exact2}) into Eq.~(\ref{eq:sin a+b=sin a con b+ cos a sin b}), and expanding the result for small $m$ with fixed $T-t_0$ and keeping the order $m$ and $m\log m$ terms results in 
\begin{align}
    \label{eq:<sin> nonperturbative piece}
    P_{1}=
     -\frac{2m}{e^{\gamma}}\sin{\sqrt{4\pi}\phi_{\rm cl}(T,Y)}\log{|me^{\gamma}(T-t_{0})|}.
\end{align}
The appearance of $m \log m$ behavior explains why a naive expansion in $m$ produced a divergence.  The hybrid method nicely separates the $m\log m$ term from the effect of the background field and allows us to obtain this result.

\subsection{Current expectation value}
We are finally ready to present the expression of the renormalized current $\braket{j^{x}}$. 
By plugging Eqs.~(\ref{eq:<sin> nonperturbative piece}) and (\ref{eq:<sin> perturbative piece}) into Eq.~(\ref{eq:<sin> expand boson}), we obtain the following expression to order $m$: 
\begin{align}
    \label{eq:<sin> expand boson smallm}
    & \Braket{\mu N_{\mu}\left\{\sin{\sqrt{4\pi}\phi(T,Y)}\right\}}
    =
     -\frac{2m}{e^{\gamma}}\sin{\sqrt{4\pi}\phi_{\rm cl}(T,Y)}\log{|me^{\gamma}(T-t_{0})|}\notag
     \\
    &\hspace{1cm}+ \frac{2m}{e^{\gamma}} \int_{t_{0}}^{T}\frac{dt_{1}}{2(T-t_{1})}\left\{
     \sin{\left[\sqrt{4\pi}\phi_{\rm cl}^{I}(T,Y)-\sqrt{4\pi}\phi_{\rm cl}^{I}(t_{1},Y-T+t_{1})\right]}\right.\notag
    \\
    &\hspace{4.5cm}\left.+
      \sin{\left[\sqrt{4\pi}\phi_{\rm cl}^{I}(T,Y)-\sqrt{4\pi}\phi_{\rm cl}^{I}(t_{1},Y+T-t_{1})\right]}
    \right\}.
\end{align}
When $T<t_{0}$, the above expression vanishes since $\phi_{\rm cl}(t,y)=0$ for $t<t_{0}$; thus, the range of time integral in Eq.~(\ref{eq:j expectation value boson}) changes from $(-\infty,t]$ to $[t_{0},t]$. 
Then, $f(T)$ in Eq.~(\ref{eq:j expectation value boson}) can be set to $f(T)=1$. 
Substituting Eq.~(\ref{eq:<sin> expand boson smallm}) into Eq.~(\ref{eq:j expectation value boson}) together with Eq.~(\ref{eq:pi phi interaction}) yields our main result:
\begin{align}
    \label{eq:j expectation value boson smallm}
    &\braket{j^{x}}= 
     -\frac{q}{\sqrt{\pi}} \partial_{t}\phi_{\rm cl}(t,x)\notag
     \\
     &
     -\frac{2m^{2}q}{\pi} \int_{t_{0}}^{t} dT 
     \frac{1}{2}\left[\sin{\sqrt{4\pi}\phi_{\rm cl}(T,x-t+T)}+
     \sin{\sqrt{4\pi}\phi_{\rm cl}(T,x+t-T)}
    \right]\log{\left|me^{\gamma}(T-t_{0})\right|}
    \notag
     \\
    &+\frac{2m^{2}q}{\pi} \int_{t_{0}}^{t} dT \int_{t_{0}}^{T}\frac{dt_{1}}{4(T-t_{1})}\left\{
     \sin{\left[\sqrt{4\pi}\phi_{\rm cl}(T,x-t+T)-\sqrt{4\pi}\phi_{\rm cl}(t_{1},x-t+t_{1})\right]}\right.\notag
    \\
    &\hspace{4.8cm}+
     \sin{\left[\sqrt{4\pi}\phi_{\rm cl}(T,x+t-T)-\sqrt{4\pi}\phi_{\rm cl}(t_{1},x+t-t_{1})\right]}\notag
    \\
    &\hspace{4.8cm}+ \sin{\left[\sqrt{4\pi}\phi_{\rm cl}(T,x-t+T)-\sqrt{4\pi}\phi_{\rm cl}(t_{1},x-t+2T-t_{1})\right]}
     \notag
    \\
    &\hspace{4.8cm}\left.+
     \sin{\left[\sqrt{4\pi}\phi_{\rm cl}(T,x+t-T)-\sqrt{4\pi}\phi_{\rm cl}(t_{1},x+t-2T+t_{1})\right]}
    \right\},
\end{align}
where $\phi_{\rm cl}$ is a solution to Eq.~(\ref{eq:phi classical}) with $\mu=0$. 
In section \ref{sec:fermi Schwinger}, we show that, when the background field is assumed to be spatially homogeneous, Eq.~(\ref{eq:j expectation value boson smallm}) agrees with the expression for the current computed purely on the fermion side using the adiabatic regularization, providing a non-trivial check of this expression.

\subsection{Regime of validity of the $m$ expansion}
\label{sec: validity of m expansion}
In order for Eq.~\eqref{eq:j expectation value boson smallm} to be a valid approximation, the first term in Eq.~\eqref{eq:j expectation value boson smallm} needs to be numerically dominant over the rest of the terms. 
To see when this happens, let $\mathcal{T}$ be the time interval over which the current is evaluated (i.e., a typical time scale of the problem $\sim t-t_0$). 
Let us denote $\Phi$ as maximum of $|\phi_{\rm cl}(t,x)|$  within a time interval $\mathcal{T}$. Then, the first term of Eq.~\eqref{eq:j expectation value boson smallm} is given by $q|\partial_{t}\phi_{\rm cl}|\gtrsim\Phi/\mathcal{T}$. The second term is $\sim qm^2\mathcal{T}\log{m\mathcal{T}}\sin{\Phi}$, and the third term is $\sim qm^2\mathcal{T}\sin{\Phi}$. Thus, the ratio of the second and third terms to the first term is given by $\lesssim(m\mathcal{T})^2(1+\log{m\mathcal{T}})\sin{\Phi}/\Phi$. Thus, as long as $m\mathcal{T}\ll 1$, the mass-correction terms (second and third terms) are perturbatively small compared to the massless contribution (first term) regardless of the size of $\Phi$. Restoring $\hbar$, this condition is written as
\begin{align}
    \label{eq: small mass condition}
    \mathcal{T}\ll \frac{\hbar}{m}.
\end{align}
In fact, we have used the condition \eqref{eq: small mass condition} explicitly in deriving Eq.~\eqref{eq:<sin> nonperturbative piece}, where we expand the modified Bessel function $K_{0}(x)$. 
Notice also that Eq.~\eqref{eq: small mass condition} does not place any restriction on the magnitude of $E_C$\footnote{
It is worth mentioning the role of the strong field condition given by $q\hbar\mathscr{E}\gg m^2$ where $\mathscr{E}$ is a typical magnitude of $E_C$. Technically, this condition is not required for the validity of Eq.~\eqref{eq:j expectation value boson smallm}. Nevertheless, the strong field condition still gives a regime in which the dependence of $\braket{j^x}$ on $E_{C}$ becomes non-linear within the time $\mathcal{T}$ satisfying Eq.~\eqref{eq: small mass condition}. 
To see this, notice that $\Phi$ is linear on $E_C$ due to Eq.~\eqref{eq:phi classical} while the mass-correction terms in Eq.~\eqref{eq:j expectation value boson smallm} depend non-linearly on $\Phi$, so the non-linearity appears when $\Phi$ reaches order unity. 
From Eq.~\eqref{eq:phi classical}, its characteristic magnitude satisfies $\Phi\sim q\mathscr{E}\mathcal{T}^2/\hbar$. Thus, $\Phi$ becomes order unity on the time scale
$\mathcal T_E\sim\sqrt{\hbar/q\mathscr E}$. 
This time scale lies within the validity regime of the mass expansion \eqref{eq: small mass condition} when the strong field condition is satisfied since $m\mathcal T_E\hbar\sim (m^2/\hbar q\mathscr E)^{1/2}\ll 1$.
Thus, in the strong field regime, $\Phi$ becomes order unity before Eq.~\eqref{eq: small mass condition} breaks down, and mass-correction terms in Eq.~\eqref{eq:j expectation value boson smallm} start to exhibit non-linear dependence on $E_C$. 
}. 

Next, we consider when  Eq.~\eqref{eq:j expectation value boson smallm} may be regarded as an approximation to the full 1+1D QED theory ~\eqref{eq: QED Hamiltonian}. The no-backreaction Hamiltonian \eqref{eq: fermion quantum hamiltonian} is obtained from 1+1D QED Hamiltonian \eqref{eq: QED Hamiltonian} in the limit of $q\to0$ while holding $qE_{C}$ and $m$ fixed. The small mass expansion was taken only after this limit was taken. A hierarchy of parameters consistent with this ordered limiting procedure is, after restoring $\hbar$, given by
\begin{align}
    \label{eq: small q compared to m}
    q\sqrt{\hbar}\ll m,
\end{align}
while $qE_C$ remains fixed. 
Combining this condition with Eq.~\eqref{eq: small mass condition}, we obtain $\mathcal{T}\ll \sqrt{\hbar}/q$. 
In 1+1D massless QED, the backreacting field associated with a constant background electric field oscillates with frequency $q/\sqrt{\pi\hbar}$ \cite{Chu.Vachaspati2010apr} (see also Eq.~(5) of paper I \cite{Gralla:2025imj}). In other words, the significant backreaction develops on the time scale of $\sqrt{\hbar}/q$. Thus, the condition \eqref{eq: small q compared to m} ensures that within the time interval $\mathcal{T}$ of the problem, the backreaction is negligible. 
Furthermore, Eq.~\eqref{eq: small q compared to m} implies that the leading backreaction contribution omitted in deriving Eq.~\eqref{eq:j expectation value boson smallm} is negligible even compared with the mass correction terms ($m^2$ and $m^2\log m$ terms) retained in Eq.~\eqref{eq:j expectation value boson smallm}. Thus, within this parameter regime, Eq.~\eqref{eq:j expectation value boson smallm} provides a good approximation to the full 1+1D QED result.

\section{Fermion calculation for a spatially homogeneous background field}\label{sec:fermi Schwinger}

As an independent check of our hybrid boson-fermion method, we now revisit the same problem using a strictly fermionic approach.  The principal difficulty of the fermion calculation is the need to regularize the coincidence limit of the operators used to define the current.  However, a practical regularization method has been developed by Ref.~\cite{Ferreiro.Navarro-Salas2018jun} for spatially homogeneous background fields $E_{C}(t)$, build on the work of \cite{Kluger.Eisenberg.ea1991oct,Kluger.Eisenberg.ea1993jun,Kluger.Mottola.ea1998nov,Cohen.McGady2008aug}.  We will apply this method perturbatively in $m$ and verify that it agrees with our result (\ref{eq:j expectation value boson smallm}) when the field is spatially homogeneous. 

Instead of the Coulomb gauge we used for the hybrid method, we use the gauge where
\begin{align}
    A^{\mu}=(0,A_{C}(t)),
\end{align}
which is possible for spatially homogeneous background fields in 1+1D.  Then, the electric field is given by 
\begin{align}
    \label{eq:Efermi}
    E_{C}(t)=-\frac{dA_{C}(t)}{dt}
    .
\end{align}
As in our hybrid calculation, we assume that $E_{C}(t)$ vanishes until time $t=t_{0}$, after which it can change arbitrarily.  

Since the theory contains no interactions (only a background field), the Heisenberg picture is a convenient way to capture its dynamics. The Dirac equation for the field operator $\widehat{\psi}^{H}(x,t)$ in the Heisenberg picture is given by 
\begin{align}
    \label{eq:Diraceq}
    \left(i\gamma^{t}\partial_{t}+i\gamma^{x}\partial_{x}+q\gamma^{x}A_{C}(t)-m\right)\widehat{\psi}^{H}(x,t)=0,
\end{align}
where the gamma matrix is given by Eq.~(\ref{eq:gamma matrix}).  The field also satisfies equal time commutation relation,
\begin{align}
    \left\{\widehat{\psi}^{H}_{\alpha}(t,x), \widehat{\psi}_{\beta}^{H\dagger}(t,y)\right\}=\delta_{\alpha\beta}\delta(x-y).
\end{align}
Using the notation in \cite{Ferreiro.Navarro-Salas2018jun}, the operator solution to Eq.~(\ref{eq:Diraceq}) can be written as 
\begin{align}
   &\widehat{\psi}^{H}(t,x)=\int_{-\infty}^{\infty}\! \frac{dk}{2\pi} \left[\widehat{b}_{k}e^{ikx}\begin{pmatrix}
        h^{I}_{k}(t)
        \\
        -h^{II}_{k}(t)
    \end{pmatrix}
    +\widehat{d}^{\dagger}_{k}e^{-ikx}
    \begin{pmatrix}
        h^{II*}_{-k}(t)
        \\
        h^{I*}_{-k}(t)
    \end{pmatrix}\right],\label{eq:psiH}
\end{align}
where $\omega_{k,m}$ is given by Eq.~(\ref{eq:frequency}), while $h^{I}_{k}(t)$ and $h^{II}_{k}(t)$ satisfy
\begin{align}
    \label{eq: ODE for modes}
    \frac{d}{dt}\begin{pmatrix}
        h^{I}_{k}(t)
        \\
        h^{II}_{k}(t)
    \end{pmatrix}
    =
    i\begin{pmatrix}
        k-qA_{C}(t) &m 
        \\
        m &-k+qA_{C}(t)
    \end{pmatrix}
    \begin{pmatrix}
        h^{I}_{k}(t)
        \\
        h^{II}_{k}(t)
    \end{pmatrix}.
\end{align}
The initial conditions for this ODE are provided by the requirement that the mode functions reduce to those of the free theory before the electric field turns on:
\begin{align}
    \label{eq:hIini}
    h^{I}_{k}(t<t_{0})=& \sqrt{\frac{\omega_{k,m}-k}{2\omega_{k,m}}}e^{-i\omega_{k,m} (t-t_{0})},
    \\
    \label{eq:hIIini}
    h^{II}_{k}(t<t_{0})=& -\sqrt{\frac{\omega_{k,m}+k}{2\omega_{k,m}}}e^{-i\omega_{k,m} (t-t_{0})},
\end{align}
This choice guarantees that the operators $\widehat{b}_{k}$ and $\widehat{d}_{k}$ appearing in \eqref{eq:psiH} are the same operators defined in Eq.~(\ref{eq:psi define b and d}) above.  In particular, the Fock vacuum $\ket{0_{\rm f},m}$ (\ref{eq: fermion vacuum}) that served as the initial state in the hybrid calculation will become the time-independent state in this Heisenberg-picture fermion calculation.  

The formulas \eqref{eq:hIini}--\eqref{eq:hIIini} also imply the useful relation
\begin{align}
    \label{eq:hIhIInorm}
    |h^{I}_{k}|^{2}+|h^{II}_{k}|^{2}=1.
\end{align} 
This equation holds by direct calculation using \eqref{eq:hIini}--\eqref{eq:hIIini} for $t>t_0$, and extends to all $t$ by the properties of the differential equation \eqref{eq: ODE for modes}.

Using adiabatic regularization, Ref.~\cite{Ferreiro.Navarro-Salas2018jun} found the following formula for the renormalized current:
\begin{align}
    \braket{j^{x}}&\equiv q\bra{0_{\rm f},m}\overline{\widehat{\psi}^{H}}\gamma^{x}\widehat{\psi}^{H}\ket{0_{\rm f},m}_{\rm{ren}}
    \label{eq:<J>fermi}
    =-\frac{q^{2}}{\pi}A_C(t)+\frac{q}{2\pi}\int^{\infty}_{-\infty}\!dk \left[|h^{I}_{k}|^{2}-|h^{II}_{k}|^{2}+\frac{k}{\omega}\right].
\end{align}
In general, one must evaluate this integral numerically using numerical solutions for $h^{I}_{k}(t)$ and $h^{II}_{k}(t)$ \cite{Pla.Newsome.ea2021may}.  However, an analytic solution is possible by expanding in $m$.  We defer the full computation to Appendix~\ref{app:derivation<j>fermi}, but comment here on a physically-interesting detail.  In evaluating the $k$-integral in Eq.~\eqref{eq:<J>fermi}  perturbatively in $m$, we find contributions from  $|k|\sim m$ that require special treatment.  Following the method of matched asymptotic expansions, we introduce an arbitrary scale $k_{c}$ that satisfies $m\ll k_{c} \ll 1/\mathcal{T}$ and split the integral into the $|k|\geq k_{c}$ and $k\leq k_{c}$ parts.  Working to order $m^2$, we find that the $|k|\leq k_{c}$ piece has an $m^2 \log m$ term, while the $|k|\geq k_{c}$ piece has only $m^2$.  Therefore, the $m^2\log m$ behavior we already found in the hybrid approach can be traced back to contributions from $k \ll m$: at least from the fermion point of view, it is an infrared effect.  The scale $k_c$ cancels out after combining the terms, and the result is
\begin{align}
    \label{eq:<j>fermismallm}
     \braket{j^{x}}
     =&
     -\frac{q^{2}}{\pi}A_{C}(t)\notag
     \\
     &+
      \frac{2qm^{2}}{\pi}\int_{t_{0}}^{t}ds
        \left[\int^{s}_{t_{0}}\frac{\sin{\left(2q\int^{s}_{w}A_{C}(t')dt'\right)}}{s-w}dw-\log{|me^{\gamma}(s-t_{0})|}\sin{\left(2q\int^{s}_{t_{0}}A_{C}(t')dt'\right)}
       \right].
\end{align}

Now, let us confirm the agreement between Eqs.~(\ref{eq:j expectation value boson smallm}) and (\ref{eq:<j>fermismallm}) for a spatially uniform electric field.  In this case, Eq.~(\ref{eq:phi classical}) becomes
\begin{align}
    \partial_{t}^{2}\phi_{\rm cl}(t)=-\frac{q}{\sqrt{\pi}}E_{C}(t).
\end{align} 
Since $\phi_{\rm cl}(t)$ satisfies this equation with the condition $\phi_{\rm cl}(t,x)=0=\partial_{t}\phi_{\rm cl}(t,x)$ for $t\leq t_{0}$, the solution is given by   
\begin{align}
    \label{eq:A to phi cl homogeneous}
   \phi_{\rm cl}(t)
   =-\frac{q}{\sqrt{\pi}}\int_{t_{0}}^{t}\int_{t_{0}}^{t'}E_{C}(s)dsdt'
    =\frac{q}{\sqrt{\pi}}\int_{t_{0}}^{t}A_{C}(t')dt',
\end{align}
where we have used Eq.~(\ref{eq:Efermi}) to write this expression in terms of the vector potential. 
Plugging Eq.~(\ref{eq:A to phi cl homogeneous}) into Eq.~(\ref{eq:j expectation value boson smallm}) and performing a straighforward computation, one can see that Eq.~(\ref{eq:j expectation value boson smallm}) agrees with Eq.~(\ref{eq:<j>fermismallm}).

\section{Discussion}\label{sec:discussion}
The main result of this paper is the analytical formula (\ref{eq:j expectation value boson smallm}) for the expectation value of the current in arbitrary time-dependent and spatially inhomogeneous fields as an expansion in the fermion mass $m$.  
Our result establishes an example of how the renormalized expectation value can be obtained in general background fields.
One point worth mentioning is the remarkable agreement of Eqs.~(\ref{eq:j expectation value boson smallm}) and (\ref{eq:<j>fermismallm}) in the spatially homogeneous background case. 
Since these two results are regularized in completely different ways (adiabatic regularization vs bosonization), our result can be thought of as an independent cross-check of the regularization method \cite{Rio.Navarro_Salas2015mar,Ferreiro.Navarro-Salas2018jun,Pla:2022spt}.

The hybrid method was crucial to our calculation, i.e, the use of the fermion-side calculation was essential in finding the non-analytic $m^2\log m$ piece in Eq.~\eqref{eq:j expectation value boson smallm}. However, if the boson--fermion duality were unknown, one would have to find this term entirely within the bosonic theory. This raises an interesting question about how the small-$m$ expansion should be treated on the bosonic side. In classical problems, non-analytic terms such $\log m$ are often incorporated by identifying a suitable solution ansatz, whereas in standard perturbative QFT, the expansion is usually assumed to be a smooth Taylor series. Therefore, it would be valuable to study the origin of the $m^2\log{m}$ piece solely within the boson theory.

The Schwinger effect in the inhomogeneous background field has been studied using the equal-time Dirac-Heisenberg-Wigner (DHW) formalism \cite{Bialynicki-Birula1991,Hebenstreit2010,Mohamedsedik2021a}, including the 1+1D case \cite{Hebenstreit:2011pm}. 
Since bosonization and DHW formalism are totally distinct approaches, it would be interesting to see how these methods give rise to the same result.

\section*{Acknowledgments}
we are grateful to Paul Anderson for many helpful conversations. This work was supported by grants
from the Simons Foundation (MPSCMPS-00001470)
and the National Science Foundation (PHY-230919).

\appendix
\section{Some properties of fermion normal-ordering}\label{app:fermi normal order}
In this appendix, we define the fermion normal-ordering and review its several properties. 
Let us mode expand $\widehat{\psi}(x)$ with an arbitrary mass $\mathfrak{m}$ as
\begin{align}
    \label{eq:psi expand arbitrary mass}
    \widehat{\psi}(x)=\begin{pmatrix}
        \widehat{\psi}_{1}(x)\\
        \widehat{\psi}_{2}(x)
    \end{pmatrix}
    =&\int_{-\infty}^{\infty}\frac{e^{-\frac{1}{2}\epsilon |k|}dk}{2\pi\sqrt{2\omega_{k,\mathfrak{m}}}}
    \left[
    \widehat{b}_{k,\mathfrak{m}}e^{ikx}\begin{pmatrix}
        \sqrt{\omega_{k,\mathfrak{m}}-k}\\
        \sqrt{\omega_{k,\mathfrak{m}}+k}
    \end{pmatrix}
    +\widehat{d}^{\dagger}_{k,\mathfrak{m}}e^{-ikx}\begin{pmatrix}
        -\sqrt{\omega_{k,\mathfrak{m}}-k}\\
        \sqrt{\omega_{k,\mathfrak{m}}+k}
    \end{pmatrix}
    \right]
\end{align}
where the factor $e^{-\frac{1}{2}\epsilon |k|}$ with $\epsilon\to0^+$ is introduced to UV-regulate the operator. 
We use $:\cdots:_{\mathfrak{m}}$ to denote the normal-ordering with respect to the mass $\mathfrak{m}$, i.e., moving all $\widehat{b}^{\dagger}_{k,\mathfrak{m}}$ and $\widehat{d}^{\dagger}_{k,\mathfrak{m}}$ to the left of all $\widehat{b}_{k,\mathfrak{m}}$ and $\widehat{d}_{k,\mathfrak{m}}$. The Fock vacuum $\ket{0_{\rm f},\mathfrak{m}}$ corresponding to the mass $\mathfrak{m}$ is defined by
\begin{align}
    \widehat{b}_{k,\mathfrak{m}}\ket{0_{\rm f},\mathfrak{m}}
    =0
    =\widehat{d}_{k,\mathfrak{m}}\ket{0_{\rm f},\mathfrak{m}}.
\end{align}
The normal-ordering  $:\cdots:_{\mathfrak{m}}$ for the quadratic operator $\widehat{\mathcal{O}}(x)$ constructed by $\widehat{\psi}(x)$ and $\widehat{\psi}^{\dagger}(x)$ can be written as 
\begin{align}
    \label{eq: normal order quadratic}
    :\widehat{\mathcal{O}}(x):_{\mathfrak{m}}=\widehat{\mathcal{O}}(x)-\bra{0_{\rm f},\mathfrak{m}}\widehat{\mathcal{O}}(x)\ket{0_{\rm f},\mathfrak{m}}.
\end{align}
Computing the expectation values of the operators $-i\widehat{\psi}(x)\gamma^{x}\partial_{x}\widehat{\psi}(x)$ and $\overline{\widehat{\psi}}(x)\widehat{\psi}(x)$ in the state $\ket{0_{\rm f},\mathfrak{m}}$, we have
\begin{align}
   \label{eq:kinetic expectation value}
   \bra{0_{\rm f},\mathfrak{m}} -i\widehat{\psi}(x)\gamma^{x}\partial_{x}\widehat{\psi}(x) \ket{0_{\rm f},\mathfrak{m}}&=-\int_{-\infty}^{\infty}\frac{e^{-\epsilon|k|}dk}{2\pi}\frac{k^2}{\omega_{k,\mathfrak{m}}},
   \\
   \label{eq:mass expectation value}
    \bra{0_{\rm f},\mathfrak{m}}
    \overline{\widehat{\psi}}(x)\widehat{\psi}(x)
    \ket{0_{\rm f},\mathfrak{m}}&=-\int_{-\infty}^{\infty}\frac{e^{-\epsilon|k|}dk}{2\pi}\frac{\mathfrak{m}}{\omega_{k,\mathfrak{m}}},
\end{align}
where the gamma matrix $\gamma^{\mu}$ $(\mu=t,x)$ is given by Eq.~\eqref{eq:gamma matrix}. 
Using Eq.~\eqref{eq: normal order quadratic}, these results imply that, for two different masses $\mathfrak{m}$ and $\mathfrak{m}'$,
\begin{align}
 : -i\widehat{\psi}(x)\gamma^{x}\partial_{x}\widehat{\psi}(x)
   :_{\mathfrak{m}}- :& -i\widehat{\psi}(x)\gamma^{x}\partial_{x}\widehat{\psi}(x)
   :_{\mathfrak{m}'}\notag
   \\
   &=\int_{-\infty}^{\infty}\frac{k^2 e^{-\epsilon|k|}dk}{2\pi}
   \left(\frac{1}{\omega_{k,\mathfrak{m}}}
   -
   \frac{1}{\omega_{k,\mathfrak{m}'}}
   \right)\notag
   \\
   &=\frac{1}{2\pi}\left[\mathfrak{m}^2\log\left(\frac{e^{\gamma+1/2}\mathfrak{m}\epsilon}{2}\right)-\mathfrak{m}'^2\log\left(\frac{e^{\gamma+1/2}\mathfrak{m}'\epsilon}{2}\right)\right]+O(\epsilon),
\end{align}
and 
\begin{align}
    :\overline{\widehat{\psi}}(x)\widehat{\psi}(x)
   :_{\mathfrak{m}}- :\overline{\widehat{\psi}}(x)\widehat{\psi}(x)
   :_{\mathfrak{m}'}
   &=\int_{-\infty}^{\infty}\frac{e^{-\epsilon|k|}dk}{2\pi}
   \left(\frac{\mathfrak{m}}{\omega_{k,\mathfrak{m}}}
   -
   \frac{\mathfrak{m}'}{\omega_{k,\mathfrak{m}'}}
   \right)\notag
   \\
   &=\frac{1}{\pi}\left[\mathfrak{m}\log\left(\frac{2}{e^{\gamma}\mathfrak{m}\epsilon}\right)-\mathfrak{m}'\log\left(\frac{2}{e^{\gamma}\mathfrak{m}'\epsilon}\right)\right]+O(\epsilon).
\end{align}
In other words, the operators 
\begin{align}
: -i\widehat{\psi}(x)\gamma^{x}\partial_{x}\widehat{\psi}(x)
   :_{\mathfrak{m}},\qquad
     :\overline{\widehat{\psi}}(x)\widehat{\psi}(x)
   :_{\mathfrak{m}}
\end{align}
depend on the reference mass $\mathfrak{m}$ used to define normal ordering, but changing $\mathfrak{m}$ to a different value only results in the additive shift of the operators by a divergent $c$-number. 

However, the operators $\overline{\widehat{\psi}}(x)\gamma^{t}\widehat{\psi}(x)$, $ \overline{\widehat{\psi}}(x)\gamma^{x}\widehat{\psi}(x)$, and $ \overline{\widehat{\psi}}(x)\gamma^{5}\widehat{\psi}(x)$ are normal-ordering mass independent. Direct calculations yield
\begin{align}
    \label{eq:charge expectation value}
    \bra{0_{\rm f},\mathfrak{m}}
    \overline{\widehat{\psi}}(x)\gamma^{t}\widehat{\psi}(x)
    \ket{0_{\rm f},\mathfrak{m}}&=\int_{-\infty}^{\infty}\frac{e^{-\epsilon |k|}dk}{2\pi}=\frac{1}{\pi \epsilon},
    \\
    \label{eq:current expectation value}
     \bra{0_{\rm f},\mathfrak{m}}
    \overline{\widehat{\psi}}(x)\gamma^{x}\widehat{\psi}(x)
    \ket{0_{\rm f},\mathfrak{m}}&=\int_{-\infty}^{\infty}\frac{e^{-\epsilon |k|}dk}{2\pi}\frac{k}{\omega_{k,\mathfrak{m}}}=0,
    \\
    \label{eq:chiral expectation value}
     \bra{0_{\rm f},\mathfrak{m}}
    \overline{\widehat{\psi}}(x)\gamma^{5}\widehat{\psi}(x)
    \ket{0_{\rm f},\mathfrak{m}}&=0.
\end{align}
Even though Eq.~\eqref{eq:charge expectation value} diverges as $\epsilon\to0^+$, it does not depend on $\mathfrak{m}$. In other words, we have
\begin{align}
     &:\overline{\widehat{\psi}}(x)\gamma^{t}\widehat{\psi}(x)
   :_{\mathfrak{m}}- :\overline{\widehat{\psi}}(x)\gamma^{t}\widehat{\psi}(x)
   :_{\mathfrak{m}'}\notag
   \\
   &\hspace{3cm}=
    \bra{0_{\rm f},\mathfrak{m}'}
    \overline{\widehat{\psi}}(x)\gamma^{t}\widehat{\psi}(x)
    \ket{0_{\rm f},\mathfrak{m}'}-\bra{0_{\rm f},\mathfrak{m}}
    \overline{\widehat{\psi}}(x)\gamma^{t}\widehat{\psi}(x)
    \ket{0_{\rm f},\mathfrak{m}}\notag
    \\
    &\hspace{3cm}=0.
\end{align}
From Eqs.~\eqref{eq:current expectation value} and \eqref{eq:chiral expectation value}, we have 
\begin{align}
     :\overline{\widehat{\psi}}(x)\gamma^{x}\widehat{\psi}(x)
   :_{\mathfrak{m}}- :\overline{\widehat{\psi}}(x)\gamma^{x}\widehat{\psi}(x)
   :_{\mathfrak{m}'}&=0.
   \\
   :\overline{\widehat{\psi}}(x)\gamma^{5}\widehat{\psi}(x)
   :_{\mathfrak{m}}- :\overline{\widehat{\psi}}(x)\gamma^{5}\widehat{\psi}(x)
   :_{\mathfrak{m}'}&=0.
\end{align}
Therefore, the operators
\begin{align}
    :\overline{\widehat{\psi}}(x)\gamma^{\mu}\widehat{\psi}(x)
   :_{\mathfrak{m}},\qquad  :\overline{\widehat{\psi}}(x)\gamma^{5}\widehat{\psi}(x)
   :_{\mathfrak{m}}
\end{align}
are normal-ordering mass $\mathfrak{m}$ \emph{independent}. 

These results show that the physical content of Eq.~\eqref{eq: QED Hamiltonian} is normal-ordering mass independent since $\overline{\widehat{\psi}}(x)\gamma^{t}\widehat{\psi}(x)=\widehat{\psi}^{\dagger}(x)\widehat{\psi}(x)$, the interaction term in Eq.~\eqref{eq: QED Hamiltonian} is normal-ordering mass independent, while the free part only shifts by a $c$-number in changing the normal-ordering mass.

\section{Interaction picture for a system with a time-varying Hamiltonian}\label{app:interactting picutre}
Since the Hamiltonian relevant to our problem has explicit time dependence, it is worth discussing the interaction picture for such cases. We mostly follow the standard approach (e.g., \cite{Peskin:1995ev,Itzykson:1980rh,Schwartz:2014sze}) but modify some definitions of the operators so that the perturbation theory becomes compatible with the explicitly time-dependent Hamiltonian. 
We begin with a Schrödinger-picture Hamiltonian $\widehat{H}_{T}(t)$, such as the one introduced in  Eq.~(\ref{eq:total Hamiltonian}).
Denoting the corresponding time evolution operator by $\widehat{U}_{T}(t,t')$, the states at two different times are related by
\begin{align}
    \ket{\psi(t)}=\widehat{U}_{T}(t,t')\ket{\psi(t')}
\end{align}
where $\widehat{U}_{T}(t,t')$ is unitary and solves the Schr\"odinger equation
\begin{align}
    \label{eq:define U(t,t')}
    i\frac{\partial}{\partial t}\widehat{U}_{T}(t,t')=&\widehat{H}_{T}(t)\widehat{U}_{T}(t,t')
\end{align}
together with $\widehat{U}_{T}(t,t)=\widehat{1}$. 
For perturbative calculations, we decompose the the Hamiltonian $\widehat{H}_{T}(t)$ into the free and the interaction parts as
\begin{align}
    \widehat{H}_{T}(t)=\widehat{H}_{0}(t)+\widehat{H}_{\rm int}(t).
\end{align}
Now, we define the free evolution operator  $\widehat{U}_{0}(t,t')$ as
\begin{align}
    \label{eq:define U0(t,t')}
    i\frac{\partial}{\partial t}\widehat{U}_{0}(t,t')=&\widehat{H}_{0}(t)\widehat{U}_{0}(t,t'),
\end{align}
with the condition $\widehat{U}_{0}(t,t)=\widehat{1}$. 
The free evolution operator $\widehat{U}_{0}(t,t')$ can then be used to transform the Schr\"odinger picture to the interaction picture.  Choosing an arbitrary reference time $t_0$, a Schr\"odinger-picture operator $\widehat{O}^{S}(t)$ (which may have explicit time dependence) becomes an interaction-picture operator$\widehat{O}^{I}(t)$ by
\begin{align}
    \label{eq:define interaction O}
    \widehat{O}^{I}(t)=\widehat{U}_{0}^{\dagger}(t,t_{0})\widehat{O}^{S}(t)\widehat{U}_{0}(t,t_{0}).
\end{align}
Differentiating Eq.~\eqref{eq:define interaction O} with respect to $t$ gives the Heisenberg equation,
\begin{align}
    \label{eq:heisenberg O interaction}
    \frac{\partial}{\partial t}\widehat{O}^{I}(t)=&\frac{1}{i}\left[\widehat{O}^{I}(t),\widehat{H}_{0}^{I}(t)\right]+\left(\frac{\partial\widehat{O}(t)}{\partial t}\right)^{I},
\end{align}
where the last term is defined by
\begin{align}
    \left(\frac{\partial\widehat{O}(t)}{\partial t}\right)^{I}\equiv 
    \widehat{U}_{0}^{\dagger}(t,t_{0})\frac{\partial \widehat{O}^{S}(t)}{\partial t}\widehat{U}_{0}(t,t_{0}).
\end{align}
The time evolution due to the interaction Hamiltonian $\widehat{H}_{\rm int}(t)$ is encoded in the operator $\widehat{U}_{\rm int}(t,t')$ defined by 
\begin{align}
    \label{eq:define Uint(t,t')}
    \widehat{U}_{\rm int}(t,t')=\widehat{U}_{0}^{\dagger}(t,t_{0})\widehat{U}_{T}(t,t')\widehat{U}_{0}(t',t_{0}),
\end{align}
which satisfies $\widehat{U}_{\rm int}(t,t)=\hat{1}$.  
Differentiating Eq.~(\ref{eq:define Uint(t,t')}) with respect to $t$ gives 
Differentiating Eq.~(\ref{eq:define Uint(t,t')}) with respect to $t$ gives 
\begin{align}
    \label{eq:schw Uint}
    i\frac{\partial}{\partial t}\widehat{U}_{\rm int}(t,t')=&\widehat{H}^{I}_{\rm int}(t)\widehat{U}_{\rm int}(t,t').
\end{align}
where we have used Eqs.~(\ref{eq:define U0(t,t')}) and (\ref{eq:heisenberg O interaction}). 
Together $\widehat{U}_{\rm int} (t,t)=\hat{1}$, Eq.~(\ref{eq:schw Uint}) has the standard time-ordered exponential
\begin{align}
    \label{eq:Uint dyson series}
    \widehat{U}_{\rm int}(t,t')=T\exp{\left(-i\int^{t}_{t'}\widehat{H}^{I}_{\rm int}(s)ds\right)}.
\end{align}

\section{Some properties of boson field operators}

In this appendix, we derive several formulas used in the main text necessary for deriving Eq.~(\ref{eq:j expectation value boson smallm}). 
It is convenient to define the positive and negative frequency parts of Eq.~(\ref{eq:op sol phi interaction}) and (\ref{eq:op sol pi interaction}) by
\begin{align}
    \label{eq:phi- interaction}
    \widehat{\phi}^{I}_{-}(t,x)=&\int_{-\infty}^{\infty}\frac{dk}{2\pi\sqrt{2\omega_{k}}}\widehat{a}_{k}e^{-i\omega_{k}(t-t_{0})+ikx-\frac{1}{2}\epsilon\omega_{k}},
    \\
    \label{eq:phi+ interaction}
    \widehat{\phi}^{I}_{+}(t,x)=&\int_{-\infty}^{\infty}\frac{dk}{2\pi\sqrt{2\omega_{k}}}\widehat{a}^{\dagger}_{k}e^{i\omega_{k}(t-t_{0})-ikx-\frac{1}{2}\epsilon\omega_{k}},
    \\
     \label{eq:pi- interaction}
    \widehat{\Pi}^{I}_{-}(t,x)=&-i\int_{-\infty}^{\infty}\frac{dk}{2\pi}\sqrt{\frac{\omega_{k}}{2}}\widehat{a}_{k}e^{-i\omega_{k}(t-t_{0})+ikx-\frac{1}{2}\epsilon\omega_{k}},
    \\
     \label{eq:pi+ interaction}
    \widehat{\Pi}^{I}_{+}(t,x)=&i\int_{-\infty}^{\infty}\frac{dk}{2\pi}\sqrt{\frac{\omega_{k}}{2}}\widehat{a}^{\dagger}_{k}e^{i\omega_{k}(t-t_{0})-ikx-\frac{1}{2}\epsilon\omega_{k}}.
\end{align}
Also, we explicitly include the UV regulator characterized by $\epsilon>0$.  
We should understand $\epsilon$ in Eqs.~(\ref{eq:phi- interaction})--(\ref{eq:pi+ interaction}) as infinitesimally small and take it to be zero at the end of the calculations.

Before proceeding further, let us clarify the procedure of the $\mu\to0$ limit. 
We keep $\mu$ finite in all operator-valued expressions and take the limit only after the relevant expression has reduced to a $c$-number quantity. This applies not only to expectation values but also to a commutator such as
\begin{align}
    \left[\widehat{\phi}^{I}_{-}(t_{1},y_{1}),\widehat{\phi}^{I}_{+}(t_{2},y_{2})\right]
\end{align}
since this expression is no longer operator valued.
Keeping this in mind, direct calculations of the commutation relations for the operators (\ref{eq:phi- interaction}) and (\ref{eq:phi+ interaction}) give 
\begin{align}
    \lim_{\mu\to0}\left[\widehat{\phi}^{I}_{-}(t_{1},y_{1}),\widehat{\phi}^{I}_{+}(t_{2},y_{2})\right]&=\lim_{\mu\to0}\int_{-\infty}^{\infty}\frac{dk}{2\pi}\frac{1}{2\omega_{k}}e^{-i\omega_{k}\Delta t+ik\Delta y-\epsilon \omega_{k}}\notag
    \\
    &=-\frac{i}{4}\lim_{\mu\to0}H^{(2)}_{0}\left(\mu\sqrt{(\Delta t-i\epsilon)^{2}-\Delta y^{2}}\right)\notag
    \\
    &\xrightarrow{\mu\to0} 
    \frac{1}{4\pi}\log{\left\{\left(-\frac{4}{e^{2\gamma}\mu^{2}}\right)\frac{1}{(\Delta t-i\epsilon)^{2}-\Delta y^{2}}\right\}}.
\end{align}
where $\Delta t=t_{1}-t_{2}$ and $\Delta y=y_{1}-y_{2}$. 
Since this result blows up as $\mu\to 0$, the commutator of $\widehat{\phi}^{I}_{-}(t_{1},y_{1})$ and $\widehat{\phi}^{I}_{+}(t_{2},y_{2})$ is not a well-defined quantity in this limit. 
However, the following expressions have the well-defined limit of $\mu\to0$:
\begin{align}
    \label{eq:e^4pi[phi-,phi+]}
    \lim_{\mu\to0}\mu^{2}e^{4\pi \left[\widehat{\phi}^{I}_{-}(t_{1},y_{1}),\widehat{\phi}^{I}_{+}(t_{2},y_{2})\right] }
    &=\left(-\frac{4}{e^{2\gamma}}\right)\frac{1}{(\Delta t-i\epsilon)^{2}-\Delta y^{2}},
    \\
    \label{eq:-e^4pi[phi-,phi+]}
     \lim_{\mu\to0}e^{-4\pi \left[\widehat{\phi}^{I}_{-}(t_{1},y_{1}),\widehat{\phi}^{I}_{+}(t_{2},y_{2})\right] }
     &= 0 .
\end{align}
Also, the commutator of $\widehat{\Pi}^{I}_{-}(t_{1},y_{1})$ and $\widehat{\phi}^{I}_{+}(t_{2},y_{2})$ has a well-defined limit of $\mu\to0$,
\begin{align}
     \label{eq:pi- phi+ interaction}
    \lim_{\mu\to0}\left[\widehat{\phi}^{I}_{-}(t_{1},y_{1}),\widehat{\Pi}^{I}_{+}(t_{2},y_{2})\right]
    &=
    \lim_{\mu\to0}\partial_{t_{2}} \left[\widehat{\phi}^{I}_{-}(t_{1},y_{1}),\widehat{\phi}^{I}_{+}(t_{2},y_{2})\right]\notag
    \\
    &= \frac{1}{2\pi}\frac{\Delta t-i\epsilon}{(\Delta t-i\epsilon)^{2}-\Delta y^{2}}.
\end{align}
In addition, the commutator of $\widehat{\Pi}^{I}(t_{1},y_{1})$ and $\widehat{\phi}^{I}(t_{2},y_{2})$ has a simple expression,
\begin{align}
     \label{eq:pi phi interaction}
    \lim_{\mu\to0}\left[\widehat{\phi}^{I}(t_{1},y_{1}),\widehat{\Pi}^{I}(t_{2},y_{2})\right]
    &=\lim_{\mu\to0}\partial_{t_{2}}\left(
    \left[\widehat{\phi}^{I}_{-}(t_{1},y_{1}),\widehat{\phi}^{I}_{+}(t_{2},y_{2})\right]
    +\left[\widehat{\phi}^{I}_{+}(t_{1},y_{1}),\widehat{\phi}^{I}_{-}(t_{2},y_{2})\right]\
    \right)\notag
    \\
    &= \frac{1}{2\pi}\frac{\Delta t-i\epsilon}{(\Delta t-i\epsilon)^{2}-\Delta y^{2}}-\frac{1}{2\pi}\frac{\Delta t+i\epsilon}{(\Delta t+i\epsilon)^{2}-\Delta y^{2}}\notag
    \\
    &=\frac{\Delta t}{2\pi}
    \left(
    \frac{1}{\Delta t^{2}-\Delta y^{2}-2i\epsilon\Delta t}
    -\frac{1}{\Delta t^{2}-\Delta y^{2}+2i\epsilon\Delta t}
    \right)+o(\epsilon^{0})\notag
    \\
    &=\frac{\Delta t}{2\pi}\times
    \mathrm{sign}(\Delta t)2i\pi\delta\left(\Delta t^{2}-\Delta y^{2}\right)\notag
    \\
    &=
    \frac{i}{2}\left[\delta(\Delta t-\Delta y)+\delta(\Delta t+\Delta y)\right].
\end{align}
To derive this result, we have used the following result:
\begin{align}
    \frac{1}{x\mp i\epsilon}=\mathrm{P.V.}\frac{1}{x}\pm i\pi\delta(x)\quad \text{for $\epsilon\to0^{+}$},
\end{align}
where P.V. denotes the principal value. Indeed, when $\Delta t=0$, Eq.~(\ref{eq:pi phi interaction}) reduces to the equal time commutation relation. 

Another useful relation is the commutator of $\mu N_{\mu}\{\cos{\sqrt{4\pi}\widehat{\phi}^{I}(t_{1},y_{2})}\}$ and  $\widehat{\Pi}^{I}(t_{2},y_{2})$. 
Note that for any operators $\widehat{A}$ and $\widehat{B}$ whose commutator is a $c$-number, we have
\begin{align}
    \left[N_{\mu}\left\{e^{\widehat{A}}\right\},\widehat{B}\right]=&\left[\widehat{A},\widehat{B}\right]N_{\mu}\left\{e^{\widehat{A}}\right\}.
\end{align}
Then,
\begin{align}
    \label{eq:cos pi commu interaction}
     &\left[\mu N_{\mu}\left\{\cos{\sqrt{4\pi}\widehat{\phi}^{I}(t_{1},y_{1})}\right\},\widehat{\Pi}^{I}(t_{2},y_{2})\right]\notag
     \\
     &\qquad=
     \frac{1}{2}\left[\mu N_{\mu}\left\{e^{i\sqrt{4\pi}\widehat{\phi}^{I}(t_{1},y_{1})}+e^{-i\sqrt{4\pi}\widehat{\phi}^{I}(t_{1},y_{1})}\right\},\widehat{\Pi}^{I}(t_{2},y_{2})\right]\notag
     \\
     &\qquad=
     i\sqrt{4\pi}\left[\widehat{\phi}^{I}(t_{1},y_{1}),\widehat{\Pi}^{I}(t_{2},y_{2})\right] \times \frac{1}{2}\mu N_{\mu}\left\{e^{i\sqrt{4\pi}\widehat{\phi}^{I}(t_{1},y_{1})}-e^{-i\sqrt{4\pi}\widehat{\phi}^{I}(t_{1},y_{1})}\right\}\notag
     \\
     &\qquad=-\sqrt{4\pi}\left[\widehat{\phi}^{I}(t_{1},y_{1}),\widehat{\Pi}^{I}(t_{2},y_{2})\right]\mu N_{\mu}\left\{\sin{\sqrt{4\pi}\widehat{\phi}^{I}(t_{1},y_{1})}\right\}. 
\end{align}
Similarly,
\begin{align}
    \label{eq:cos phi commu interaction}
    &\left[\mu N_{\mu}\left\{\cos{\sqrt{4\pi}\widehat{\phi}^{I}(t_{1},y_{1})}\right\},\widehat{\phi}^{I}(t_{2},y_{2})\right]
     \notag
     \\
     &\qquad=-\sqrt{4\pi}\left[\widehat{\phi}^{I}(t_{1},y_{1}),\widehat{\phi}^{I}(t_{2},y_{2})\right]\mu N_{\mu}\left\{\sin{\sqrt{4\pi}\widehat{\phi}^{I}(t_{1},y_{1})}\right\}. 
\end{align}
The final piece needed for the main text is the commutator of $\mu N_{\mu}\{\cos{\sqrt{4\pi}\widehat{\phi}^{I}(t_{1},y_{1})}\}$ and $\mu N_{\mu}\{\sin{\sqrt{4\pi}\widehat{\phi}^{I}(t_{1},y_{1})}\}$. 
To simplify the notation, let us define two operators $\widehat{A}(t,y)$ and $\widehat{B}(t,y)$ by
\begin{align}
    \label{eq;define A}
    \widehat{A}(t,y)&= \frac{\mu}{2}N_{\mu} \left\{e^{i\sqrt{4\pi}\widehat{\phi}^{I}(t,y)}\right\}=\frac{\mu}{2} e^{i\sqrt{4\pi}\widehat{\phi}_{+}^{I}(t,y)}e^{i\sqrt{4\pi}\widehat{\phi}_{-}^{I}(t,y)}e^{i\sqrt{4\pi}\phi_{\rm cl}(t,y)},
    \\
    \label{eq:define B}
    \widehat{B}(t,y)&= \frac{\mu}{2}N_{\mu} \left\{e^{-i\sqrt{4\pi}\widehat{\phi}^{I}(t,y)}\right\}=\frac{\mu}{2} e^{-i\sqrt{4\pi}\widehat{\phi}_{+}^{I}(t,y)}e^{-i\sqrt{4\pi}\widehat{\phi}_{-}^{I}(t,y)}e^{-i\sqrt{4\pi}\phi_{\rm cl}(t,y)}.
\end{align}
With this definition, $\widehat{A}(t,y)$ and $\widehat{B}(t,y)$ are $\mu$ independent. 
Then, 
$\mu N_{\mu}\{\cos{\sqrt{4\pi}\widehat{\phi}^{I}(t,y)}\}$ and $\mu N_{\mu}\{\sin{\sqrt{4\pi}\widehat{\phi}^{I}(t,y)}\}$ in terms of $\widehat{A}(t,y)$ and $\widehat{B}(t,s)$ are given by
\begin{align}
    \label{eq:cos=A+B}
    \mu N_{\mu}\left\{\cos{\sqrt{4\pi}\widehat{\phi}^{I}(t},y)\right\}
    &=\widehat{A}(t,y)+\widehat{B}(t,y),
    \\
    \label{eq:sin=A-B}
    \mu N_{\mu}\left\{\sin{\sqrt{4\pi}\widehat{\phi}^{I}(t},y)\right\}
    &=\frac{1}{i}\left(\widehat{A}(t,y)-\widehat{B}(t,y)\right).
\end{align}
Then using the BCH formula
\begin{align}
    \label{eq:BHCformula}
    e^{\widehat{A}}e^{\widehat{B}}=e^{\widehat{A}+\widehat{B}}e^{\frac{1}{2}\left[\widehat{A},\widehat{B}\right]}=e^{\widehat{B}}e^{\widehat{A}}e^{\left[\widehat{A},\widehat{B}\right]}. 
\end{align}
for operators $\widehat{A}$ and $\widehat{B}$ whose commutator is a $c$-number, and (\ref{eq:e^4pi[phi-,phi+]}), and (\ref{eq:-e^4pi[phi-,phi+]}), we have
\begin{align}
    \label{eq:AB=BA=finite}
    \lim_{\mu\to0}\bra{0,\mu}\widehat{A}(t_{1},y_{1})\widehat{B}(t_{2},y_{2})\ket{0,\mu}
    &
    =
   \lim_{\mu\to0}  \left(-\frac{4}{e^{2\gamma}\mu^{2}}\right)
    \frac{\bra{0,\mu}N_{\mu}\left\{\widehat{A}(t_{1},y_{1})\widehat{B}(t_{2},y_{2})\right\}\ket{0,\mu}}{(\Delta t-i\epsilon)^{2}-\Delta y^{2}},\notag
    \\
    &=-\frac{1}{e^{2\gamma}}
    \frac{e^{i\sqrt{4\pi}(\phi_{\rm cl}(t_{1},y_{1})-\phi_{\rm cl}(t_{2},y_{2}))}}{(\Delta t-i\epsilon)^{2}-\Delta y^{2}},
\end{align}
and similarly
\begin{align}
    \label{eq:AB=BA=finitev2}
    \lim_{\mu\to0}\bra{0,\mu}\widehat{B}(t_{1},y_{1})\widehat{A}(t_{2},y_{2})\ket{0,\mu}
    &=
    -\frac{1}{e^{2\gamma}}
    \frac{e^{-i\sqrt{4\pi}(\phi_{\rm cl}(t_{1},y_{1})-\phi_{\rm cl}(t_{2},y_{2}))}}{(\Delta t-i\epsilon)^{2}-\Delta y^{2}},
\end{align}
where the $\phi_{\rm cl}(t,x)$ solves Eq.~(\ref{eq:phi classical}) with $\mu=0$; thus, $\phi_{\rm cl}(t.x)$ is a solution to a wave equation with a source $-qE_{C}/\sqrt{\pi}$.
Also,
\begin{align}
    \label{eq:AA=BB=0}
    & \phantom{....}\lim_{\mu\to0}\bra{0,\mu}\widehat{A}(t_{1},y_{1})\widehat{A}(t_{2},y_{2})\ket{0,\mu}=0= \lim_{\mu\to0}\bra{0,\mu}\widehat{B}(t_{1},y_{1})\widehat{B}(t_{2},y_{2})\ket{0,\mu}.
\end{align}
Equations~(\ref{eq:cos=A+B}) and (\ref{eq:sin=A-B}) together with (\ref{eq:AB=BA=finite})--(\ref{eq:AA=BB=0}), we obtain
\begin{align}
    \label{eq:cos sin commu}
    &
    \lim_{\mu\to0}\bra{0,\mu}\left[\mu N_{\mu}\left\{\cos{\sqrt{4\pi}\widehat{\phi}^{I}(t_{1},y_{1})}\right\},\mu N_{\mu}\left\{\sin{\sqrt{4\pi}\widehat{\phi}^{I}(t_{2},y_{2})}\right\}\right]\ket{0,\mu}\notag
    \\
    &\qquad=
    \lim_{\mu\to0}
    \left(-\frac{1}{i}\bra{0,\mu}\left[\widehat{A}(t_{1},y_{1}),\widehat{B}(t_{2},y_{2})\right]\ket{0,\mu}
    +\frac{1}{i}\bra{0,\mu}\left[\widehat{B}(t_{1},y_{1}),\widehat{A}(t_{2},y_{2})\right]\ket{0,\mu}\right)\notag
    \\
    &\qquad=
    \frac{1}{i} \lim_{\mu\to0}\left[
    -\bra{0,\mu}\widehat{A}(t_{1},y_{1})\widehat{B}(t_{2},y_{2})\ket{0,\mu}+\bra{0,\mu}\widehat{B}(t_{1},y_{1})\widehat{A}(t_{2},y_{2})\ket{0,\mu}\right.
    \notag
    \\
    &\hspace{5cm}\left.-\bra{0,\mu}\widehat{A}(t_{1},y_{1})\widehat{B}(t_{2},y_{2})\ket{0,\mu}^*+\bra{0,\mu}\widehat{B}(t_{1},y_{1})\widehat{A}(t_{2},y_{2})\ket{0,\mu}^*
    \right]
    \notag
    \\
    &\qquad=
    \frac{2}{e^{2\gamma}}\sin{\left[\sqrt{4\pi}\phi_{\rm cl}(t_{1},y_{1})-\sqrt{4\pi}\phi_{\rm cl}(t_{2},y_{2})\right]}\left(
    \frac{1}{(\Delta t-i\epsilon)^{2}-\Delta y^{2}}
    -\frac{1}{(\Delta t+i\epsilon)^{2}-\Delta y^{2}}
    \right)
    \notag
   \\
    &\qquad=i\left(\frac{4\pi}{e^{2\gamma}}\right)\mathrm{sign}(\Delta t)\delta\left(\Delta t^{2} -\Delta y^{2}\right) 
    \sin{\left[\sqrt{4\pi}\phi_{\rm cl}(t_{1},y_{1})-\sqrt{4\pi}\phi_{\rm cl}(t_{2},y_{2})\right]}.
\end{align}
From the first equality to the second, we have used $\widehat{A}^{\dagger}(t,y)=\widehat{B}(t,y)$ to write
\begin{align}
    \bra{0,\mu}\widehat{A}(t_{2},y_{2})\widehat{B}(t_{1},y_{1})\ket{0,\mu}=\bra{0,\mu}\widehat{A}(t_{1},y_{1})\widehat{B}(t_{2},y_{2})\ket{0,\mu}^*,
    \\
    \bra{0,\mu}\widehat{B}(t_{2},y_{2})\widehat{A}(t_{1},y_{1})\ket{0,\mu}=\bra{0,\mu}\widehat{B}(t_{1},y_{1})\widehat{A}(t_{2},y_{2})\ket{0,\mu}^*,
\end{align}
where $*$ denotes the complex conjugate.

\section{Derivation of \texorpdfstring{$\braket{j^{x}}$}{Lg} in the small mass limit  on the fermion side}\label{app:derivation<j>fermi}
To find the small mass limit \eqref{eq: small mass condition} of the current expectation value \eqref{eq:<J>fermi}, our strategy is to solve Eq.~(\ref{eq: ODE for modes}) perturbatively in $m$ and to use the perturbative solution to evaluate Eq.~\eqref{eq:<J>fermi}. 
One subtlety is that the momentum integral that appears in Eq.~(\ref{eq:<J>fermi}) has a large contribution from the region of  $k\sim m$. 
To deal with this, we consider two regions $k\gg m$ and $k\sim m$ and use the matched asymptotic approximation.

First, consider $k\gg m$ case. 
At $t<t_{0}$, no electric field exits, i.e. $A_C(t)=0$; thus the mode functions have to reduce to the free theory mode functions given by Eqs.~(\ref{eq:hIini}) and (\ref{eq:hIIini}). 
In the small mass limit with $k\gg m$ and $t<t_{0}$, $h^{I}_{k}(t)$ and $h^{II}_{k}(t)$ can be expanded as follows:
\begin{align}
    h^{I}_{k}(t)=\theta(-k)\left[
    1+\frac{-1+4ik(t-t_{0})}{8}\left(\frac{m^{2}}{k^{2}}\right)\right]e^{ik(t-t_{0})}
    +\theta(k)\frac{m}{2k}e^{-ik(t-t_{0})}+\mathcal{O}\left(\frac{m^{3}}{k^{3}}\right),
    \\
    h^{II}_{k}(t)=\theta(k)\left[
    -1+\frac{1+4ik(t-t_{0})}{8}\left(\frac{m^{2}}{k^{2}}\right)\right]e^{-ik(t-t_{0})}
    +\theta(-k)\frac{m}{2k}e^{ik(t-t_{0})}+\mathcal{O}\left(\frac{m^{3}}{k^{3}}\right)
\end{align}
where $\theta(k)$ is a step function. 
For an arbitrary time $t$, we also work to second order in $m$ since the non-trivial contribution to $\braket{j^{x}}$ appears at this order. 
Expanding $h^{I}_{k}(t)$ and $h^{II}_{k}(t)$ , we have
\begin{align}
    \label{eq:hIexpand}
    h^{I}_{k}(t)=&x^{(0)}(t)+x^{(1)}(t)+x^{(2)}(t)+\mathcal{O}\left(\frac{m^{3}}{k^{3}}\right),
    \\
    \label{eq:hIIexpand}
     h^{II}_{k}(t)=&y^{(0)}(t)+y^{(1)}(t)+y^{(2)}(t)+\mathcal{O}\left(\frac{m^{3}}{k^{3}}\right),
\end{align}
where $x^{(n)}$ and $y^{(n)}$ are order $(m/k)^{n}$. 
At $t<t_{0}$, Eqs.~(\ref{eq:hIexpand}) and (\ref{eq:hIIexpand}) have to reduce to Eqs.~(\ref{eq:hIini}) and (\ref{eq:hIIini}); thus, we have the following conditions at $t<t_{0}$:
\begin{align}
    \label{eq:x0ini}
    x^{(0)}(t)=&\theta(-k)e^{ik(t-t_{0})},
    \\
    \label{eq:y0ini}
    y^{(0)}(t)=&-\theta(k)e^{-ik(t-t_{0})},
    \\
    \label{eq:x1ini}
    x^{(1)}(t)=&\theta(k)\frac{m}{2k}e^{-ik(t-t_{0})},
    \\
    \label{eq:y1ini}
    y^{(1)}(t)=&\theta(-k)\frac{m}{2k}e^{ik(t-t_{0})},
    \\
    \label{eq:x2ini}
    x^{(2)}(t)=&\theta(-k)\left(-\frac{1}{8}+\frac{ik(t-t_{0})}{2}\right)\frac{m^{2}}{k^{2}}e^{ik(t-t_{0})},
    \\
    \label{eq:y2ini}
    y^{(2)}(t)=&\theta(k)\left(\frac{1}{8}+\frac{ik(t-t_{0})}{2}\right)\frac{m^{2}}{k^{2}}e^{-ik(t-t_{0})}.
\end{align}
By plugging Eqs.~(\ref{eq:hIexpand}) and (\ref{eq:hIIexpand}) into Eq.~(\ref{eq: ODE for modes}) and collecting terms order by order, we obtain the hierarchy of equations. 
From the zeroth order, we have
\begin{align}
    \label{eq:x0ODE}
    \dot{x}^{(0)}=&i(k-qA_{C}(t))x^{(0)},
    \\
    \label{eq:y0ODE}
    \dot{y}^{(0)}=&-i(k-qA_{C}(t))y^{(0)}.
\end{align}
For the first order,
\begin{align}
    \label{eq:x1ODE}
    \dot{x}^{(1)}=&i(k-qA_{C}(t))x^{(1)}+i m y^{(0)},
    \\
    \label{eq:y1ODE}
    \dot{y}^{(1)}=&-i(k-qA_{C}(t))y^{(1)} +i m x^{(0)},
\end{align}
and the second order is 
\begin{align}
    \label{eq:x2ODE}
    \dot{x}^{(2)}=&i(k-qA_{C}(t))x^{(2)}+i m y^{(1)},
    \\
    \label{eq:y2ODE}
    \dot{y}^{(2)}=&-i(k-qA_{C}(t))y^{(2)} +i m x^{(1)}.
\end{align}
We can solve Eqs.~(\ref{eq:x0ODE}) and (\ref{eq:y0ODE}) immediately and find 
\begin{align}
    \label{eq:x0sol}
    x^{(0)}(t)=&\theta(-k)e^{i\chi(t,t_{0})}
    \\
    \label{eq:y0sol}
     y^{(0)}(t)=&-\theta(k)e^{-i\chi(t,t_{0})}
\end{align}
where $\chi(t,t_{0})$ is defined as
\begin{align}
    \label{eq:definechi}
    \chi(t,t_{0})=\int_{t_{0}}^{t}(k-qA(t'))dt'.
\end{align}
It is easy to check that Eqs.~(\ref{eq:x0sol}) and (\ref{eq:y0sol}) indeed satisfy Eqs.~(\ref{eq:x0ini}) and (\ref{eq:y0ini}) at $t<t_{0}$, respectively. 
By elementary calculations, the solution to Eqs.~(\ref{eq:x1ODE}) and (\ref{eq:y1ODE}) satisfying Eqs.~(\ref{eq:x1ini}) and (\ref{eq:y1ini}) are given by
\begin{align}
    \label{eq:x1sol}
    x^{(1)}(t)=&-im\theta(k)e^{i\chi(t,t_{0})}\left(\int^{t}_{t_{0}}e^{-2i\chi(s,t_{0})}ds-\frac{1}{2ik}\right),
    \\
    \label{eq:y1sol}
    y^{(1)}(t)=&im\theta(-k)e^{-i\chi(t,t_{0})}\left(\int^{t}_{t_{0}}e^{2i\chi(s,t_{0})}ds+\frac{1}{2ik}\right).
\end{align}
Similarly, the solution to Eqs.~(\ref{eq:x2ODE}) and (\ref{eq:y2ODE}) satisfying Eqs.~(\ref{eq:x2ini}) and (\ref{eq:y2ini}) are found to be
\begin{align}
    \label{eq:x2sol}
    x^{(2)}(t)=&m^{2}\theta(-k)e^{i\chi(t,t_{0})}\left[
    -\int_{t_{0}}^{t}e^{-2i\chi(s,t_{0})}\left(\int^{s}_{t_{0}} e^{2i\chi(w,t_{0})}dw+\frac{1}{2ik}\right)ds
    -\frac{1}{8k^{2}}
    \right],
    \\
    \label{eq:y2sol}
    y^{(2)}(t)=&m^{2}\theta(k)e^{-i\chi(t,t_{0})}\left[
    \int_{t_{0}}^{t}e^{2i\chi(s,t_{0})}\left(\int^{s}_{t_{0}} e^{-2i\chi(w,t_{0})}dw-\frac{1}{2ik}\right)ds
    +\frac{1}{8k^{2}}
    \right].
\end{align}
Notice that $\theta(k)$ is multiplied by $x^{(1)},y^{(0)}$, and $y^{(2)}$, and $\theta(-k)$ is multiplied by $x^{(0)},x^{(2)}$. and $y^{(1)}$. 
This structure continues to higher order terms: the odd order of $x^{(n)}$ and even order of $y^{(n)}$ always get multiplied by $\theta(k)$, and the even order of $x^{(n)}$ and the odd order of $y^{(n)}$ always get multiplied by $\theta(-k)$. 
In other words, $h^{I}_{k}$ and $h^{II}_{k}$ are always written as
\begin{align}
    \label{eq:hIexpand2}
    h^{I}_{k}=&\theta(k)\left(x^{(1)}+x^{(3)}+\cdots\right)+\theta(-k)\left(x^{(0)}+x^{(2)}+\cdots\right),
    \\
    \label{eq:hIIexpand2}
    h^{II}_{k}=&\theta(-k)\left(y^{(1)}+y^{(3)}+\cdots\right)+\theta(k)\left(y^{(0)}+y^{(2)}+\cdots\right).
\end{align}
The important consequence of this structure is that $|h^{I}_{k}|^{2}$ and $|h^{II}_{k}|^{2}$ only contain the even order terms in $m$. 
The odd terms all vanish due to the property of the step fucntion, namely  $\theta(k)\theta(-k)=0$. 

Next, we consider the $k\sim m$ case. 
In this case, since $k$ is in the same order as $m$, we need to consider a different expansion than the ones used above, which can be done by considering $m \mathcal{T}$ as the expansion parameter, where $\mathcal{T}$ is a typical time scale of the problem ($t-t_0\sim\mathcal{T}$) discussed in Sec.~\ref{sec: validity of m expansion}.  
Before the electric field is turned on ($t<t_{0}$), the free field mode functions (\ref{eq:hIini}) and (\ref{eq:hIIini}) can now be expanded as
\begin{align}
    \label{hIinismallk}
    h^{I}_{k}(t)=&\sqrt{\frac{\omega-k}{2\omega}}-i\sqrt{\frac{\omega(\omega-k)}{2}}(t-t_{0})+\mathcal{O}(m^{2}\mathcal{T}^{2}),
    \\
    \label{eq:hIIinismallk}
    h^{II}_{k}(t)=&-\sqrt{\frac{\omega+k}{2\omega}}+i\sqrt{\frac{\omega(\omega+k)}{2}}(t-t_{0})+\mathcal{O}(m^{2}\mathcal{T}^{2}).
\end{align}
Here, we write $\omega\equiv\omega_{k,m}$ to simplify the expressions. 
Similar to the $k\gg m$ case, let us expand $h^{I}_{k}(t)$ and $h^{II}_{k}(t)$ in $m$ with $k\sim m$ as
\begin{align}
    \label{eq:hIexpandsmallk}
    h^{I}_{k}(t)=&X^{(0)}(t)+X^{(1)}(t)+\mathcal{O}(m^{2}\mathcal{T}^{2}),
    \\
    \label{eq:hIIexpandsmallk}
    h^{II}_{k}(t)=&Y^{(0)}(t)+Y^{(1)}(t)+\mathcal{O}(m^{2}\mathcal{T}^{2}).
\end{align}
Since $h^{I}_{k}(t)$ and $h^{II}_{k}(t)$ satisfy Eqs.~(\ref{hIinismallk}) and (\ref{eq:hIIinismallk}), we find the following conditions
\begin{align}
    \label{eq:x0inismallk}
    X^{(0)}(t)=&\sqrt{\frac{\omega-k}{2\omega}},
    \\
    \label{eq:y0inismallk}
    Y^{(0)}(t)=&-\sqrt{\frac{\omega+k}{2\omega}},
    \\
    \label{eq:x1inismallk}
    X^{(1)}(t)=&-i\sqrt{\frac{\omega(\omega-k)}{2}}(t-t_{0}),
    \\
    \label{eq:y1inismallk}
    Y^{(1)}(t)=&i\sqrt{\frac{\omega(\omega+k)}{2}}(t-t_{0}).
\end{align}
By plugging Eqs.~(\ref{eq:hIexpandsmallk}) and (\ref{eq:hIIexpandsmallk}) into Eq.~(\ref{eq: ODE for modes}), we obtain the following equations for the zeroth order
\begin{align}
    \label{eq:x0ODEsmallk}
    \dot{X}^{(0)}=&-iqA_{C}(t)X^{(0)},
    \\
    \label{eq:y0ODEsmallk}
    \dot{Y}^{(0)}=&+iqA_{C}(t)Y^{(0)},
\end{align}
and for the first order
\begin{align}
    \label{eq:x1ODEsmallk}
    \dot{X}^{(1)}=&-iqA_{C}(t)X^{(1)}+ikX^{(0)}+i m Y^{(0)},
    \\
    \label{eq:y1ODEsmallk}
    \dot{Y}^{(1)}=&+iqA_{C}(t)Y^{(1)}-ikY^{(0)} +i m X^{(0)}.
\end{align}
The solutions to Eqs.~(\ref{eq:x0ODEsmallk})--(\ref{eq:y1ODEsmallk}) satisfying Eqs.~(\ref{eq:x0inismallk})--(\ref{eq:y1inismallk}) are given by 
\begin{align}
    \label{eq:x0solsmallk}
    X^{(0)}(t)=&\sqrt{\frac{\omega-k}{2\omega}}e^{i\Psi(t,t_{0})},
    \\
    \label{eq:y0solsmallk}
    Y^{(0)}(t)=&-\sqrt{\frac{\omega+k}{2\omega}}e^{-i\Psi(t,t_{0})},
    \\
    \label{eq:x1solsmallk}
    X^{(1)}(t)=&ie^{i\Psi(t,t_{0})}\int^{t}_{t_{0}}\left(k\sqrt{\frac{\omega-k}{2\omega}}-m\sqrt{\frac{\omega+k}{2\omega}}e^{-2i\Psi(s,t_{0})}\right)ds,
    \\
    \label{eq:y1solsmallk}
    Y^{(1)}(t)=&ie^{-i\Psi(t,t_{0})}\int^{t}_{t_{0}}\left(k\sqrt{\frac{\omega+k}{2\omega}}+m\sqrt{\frac{\omega-k}{2\omega}}e^{2i\Psi(s,t_{0})}\right)ds,
\end{align}
where $\Psi(t,t_{0})$ is defined as
\begin{align}
    \label{eq:definePsi}
    \Psi(t,t_{0})=-q\int^{t}_{t_{0}}A(t')dt'.
\end{align}
It is straightforward to check that Eqs.~(\ref{eq:x0solsmallk}) and (\ref{eq:y0solsmallk}) reduce to Eqs.~(\ref{eq:x0inismallk}) and (\ref{eq:y0inismallk}) at $t<t_{0}$. 
To confirm that Eqs.~(\ref{eq:x1solsmallk}) and (\ref{eq:y1solsmallk}) indeed reduce to Eqs.~(\ref{eq:x1inismallk}) and (\ref{eq:y1inismallk}), note that the following identities hold:
\begin{align}
    -\sqrt{\frac{\omega(\omega-k)}{2}}=&k\sqrt{\frac{\omega-k}{2\omega}}-m\sqrt{\frac{\omega+k}{2\omega}},
    \\
    \sqrt{\frac{\omega(\omega+k)}{2}}=&m\sqrt{\frac{\omega-k}{2\omega}}+k\sqrt{\frac{\omega+k}{2\omega}}.
\end{align}
These identities can be verrifed by taking the square of both sides. 
Then, it is again straightforward that Eqs.~(\ref{eq:x1solsmallk}) and (\ref{eq:y1solsmallk}) become Eqs.~(\ref{eq:x1inismallk}) and (\ref{eq:y1inismallk}) at $t<t_{0}$. 

Now, with the perturbative expressions for the mode functions (\ref{eq:x0sol})--(\ref{eq:y2sol}) for $k\gg  m$ and (\ref{eq:x0solsmallk})--(\ref{eq:y1sol}) for $k\sim m$, let us compute the current expectation value (\ref{eq:<J>fermi}) in the small mass limit. 
We introduce a cutoff scale $k_{c}$ that is much greater than $m$ but still $k_{c}\to0$ in the $m\to 0$ limit. 
One example of setting $k_{c}$ is $k_{c}=(m/\mathcal{T}^{p-1})^{1/p}$ for an integer  $p>1$. Noting that $m\mathcal{T}\ll1$ holds, this gives the desired cutoff scale since $k_{c}\gg m$ is always satisfied for $m\mathcal{T}\ll 1$ and we also have $k_{c}\to 0$ as $m\to0$.
The second term of Eq.~(\ref{eq:<J>fermi}) has a huge contribution from the $k\sim m$ region; thus, we separate the second term of Eq.~(\ref{eq:<J>fermi}) into two parts so that one part concerns the region of $k\sim m$ while the other part deals with the $k\gg m$ region:
\begin{align}
    &\frac{q}{2\pi}\int^{\infty}_{-\infty}dk \left[|h^{I}_{k}|^{2}-|h^{II}_{k}|^{2}+\frac{k}{\omega}\right]=\frac{q}{2\pi}\left(\int^{\infty}_{k_{c}}+\int^{-k_{c}}_{-\infty}+\int^{k_{c}}_{-k_{c}}\right)dk \left[|h^{I}_{k}|^{2}-|h^{II}_{k}|^{2}+\frac{k}{\omega}\right].
\end{align}
The condition $k\gg m$ is always satisfied for the first two terms of the RHS of the above expression; thus, we can use the perturbed mode functions (\ref{eq:x0sol})--(\ref{eq:y2sol}). For the third term, $k$ can be the same order as $m$, so we need to use the mode functions (\ref{eq:x0solsmallk})--(\ref{eq:y1solsmallk}). 

We first consider the integral for the $k\gg m$ region. 
Note that $|h^{I}_{k}|^{2}$ and $|h^{II}_{k}|^{2}$ only contain the even order terms in $m$ due to Eqs.~(\ref{eq:hIexpand2}) and (\ref{eq:hIIexpand2}). 
In addition, from Eq.~(\ref{eq:hIhIInorm}), we have
\begin{align}
    1&=|h^{I}_{k}|^{2}+|h^{II}_{k}|^{2}\notag
    \\
    &=|x^{(0)}|^{2}+|y^{(0)}|^{2}+2\mathrm{Re}\left(x^{(0)*}x^{(2)}+y^{(0)*}y^{(2)}\right)+|x^{(1)}|^{2}+|y^{(1)}|^{2}+\mathcal{O}\left(\frac{m^{4}}{k^{4}}\right).
\end{align}
From Eqs.~(\ref{eq:x0sol}) and (\ref{eq:y0sol}), we find that $|x^{(0)}|^{2}+|y^{(0)}|^{2}=1$ holds. 
Also, notice that $x^{(0)}$, $x^{(0)}$, and $y^{(1)}$ are multiplied by $\theta(-k)$ while $y^{(0)}$, $y^{(2)}$, and $x^{(1)}$ are multiplied by $\theta(k)$. 
This implies
\begin{align}
    \label{eq:xnorm2nd}
    2\mathrm{Re}\left(x^{(0)*}x^{(2)}\right)+|y^{(1)}|^{2}=&0,
    \\
    \label{eq:ynorm2nd}
    2\mathrm{Re}\left(y^{(0)*}y^{(2)}\right)+|x^{(1)}|^{2}=&0.
\end{align}
Then, the large $k$ part of the integral becomes a straightforward calculation, which we present as follows. 
Using Eqs.~(\ref{eq:hIexpand}) and (\ref{eq:hIIexpand}), we have
\begin{align}
    &\frac{q}{2\pi}\left(\int^{\infty}_{k_{c}}+\int^{-k_{c}}_{-\infty}\right)dk \left[|h^{I}_{k}|^{2}-|h^{II}_{k}|^{2}+\frac{k}{\omega}\right]\notag
    \\
    =&
    \label{eq:order correction integrand}
     \frac{q}{2\pi}\left(\int^{\infty}_{k_{c}}+\int^{-k_{c}}_{-\infty}\right)dk 
     \left[
     4\mathrm{Re}\left(x^{(0)*}x^{(2)}\right)-4\mathrm{Re}\left(y^{(0)*}y^{(2)}\right)-\frac{k}{|k|}\frac{m^{2}}{2k^{2}}
    +\mathcal{O}\left(\frac{m^{4}}{k^{4}}\right) \right]
\end{align}
To obtain this, we used Eqs.~(\ref{eq:xnorm2nd}) and (\ref{eq:ynorm2nd}). 
Then, using Eqs.~(\ref{eq:x0sol}), (\ref{eq:y0sol}), (\ref{eq:x2sol}), and (\ref{eq:y2sol}), 
\begin{align}
     =&
      \frac{qm^{2}}{2\pi}\left(\int^{\infty}_{k_{c}}+\int^{-k_{c}}_{-\infty}\right)dk
      \left\{
      4\theta(-k)\mathrm{Re}\left[
      -\int^{t}_{t_{0}}ds\int^{s}_{t_{0}}dw e^{-2i\chi(s,w)}
      -\int^{t}_{t_{0}}ds\frac{e^{-2i\chi(s,t_{0})}}{2ik}-\frac{1}{8k^{2}}
      \right]\right.\notag
      \\
      &\left.
     \hspace{3cm} -4\theta(k)\mathrm{Re}\left[
       -\int^{t}_{t_{0}}ds\int^{s}_{t_{0}}dw e^{2i\chi(s,w)}
      +\int^{t}_{t_{0}}ds\frac{e^{2i\chi(s,t_{0})}}{2ik}-\frac{1}{8k^{2}}
      \right]
       -\frac{k}{|k|}\frac{1}{2k^{2}}
      \right\}\notag
      \\
      =&
      \label{eq:long calc}
       \frac{2qm^{2}}{\pi}\left(\int^{\infty}_{k_{c}}-\int^{-k_{c}}_{-\infty}\right)dk
       \left[
        \int^{t}_{t_{0}}ds\int^{s}_{t_{0}}dw \cos{\left(2\chi(s,w)\right)}
      -\int^{t}_{t_{0}}ds\frac{\sin{\left(2\chi(s,t_{0})\right)}}{2k}
       \right],
\end{align}
From the first equality to the second, the terms that are independent of $t$ are all canceled with each other. 
The expression Eq.~\eqref{eq:long calc} has a correction of order $\mathcal{O}(m^{4}/k_{c}^{3})$. This is due to the fact that since the correction to the integrand of Eq.~\eqref{eq:order correction integrand} is $\mathcal{O}(m^4/k^4)$, the $k$ integral bounded by $k_c$ produces the correction of order $\mathcal{O}(m^{4}/k_{c}^{3})$. 
Since $\chi(t,t_{0})$ defined in Eq.~(\ref{eq:definechi}) can be written using $\Psi(t,t_{0})$ given in Eq.~(\ref{eq:definePsi}) as $\chi(t,t_{0})=k(t-t_{0})+\Psi(t,t_{0})$, we can expand $\cos$ and $\sin$, and further carry out calculations using the fact that the odd part in $k$ gives nonzero result while the even part becomes identically zero due to symmetry of integral,
\begin{align}
    =&
    \frac{2qm^{2}}{\pi}\int_{t_{0}}^{t}ds\int^{\infty}_{k_{c}}dk
        \left[\int^{s}_{t_{0}} -2\sin{2k(s-w)}\sin{2\Psi(s,w)}dw-\frac{\cos{2k(s-t_{0})}\sin{2\Psi(s,t_{0})}}{k}
       \right].
\end{align}
Now, the $k$ integral can be done
\begin{align}
    \label{eq:kinteg1}
    &-\int_{k_{c}}^{\infty}2\sin{2k(s-w)}dk=-\frac{1}{s-w}\left(1+\mathcal{O}(k_{c}\mathcal{T})\right),
    \\
    \label{eq:kinteg2}
    &-\int_{k_{c}}^{\infty}\frac{\cos{2k(s-t_{0})}}{k}dk=\log{|2k_{c}e^{\gamma}(s-t_{0})|}+\mathcal{O}(k_{c}\mathcal{T}),
\end{align}
where $\gamma$ is the Euler constant. 
The integral that appears on the LHS of Eq.~(\ref{eq:kinteg1}) converges to the RHS within the prescription that introduces an infinitesimal positive number $\epsilon>0$ and consider a cutoff as $dk\to e^{-\epsilon k}dk$. 
This prescription gives a finite result for the integral and also does not depend on the details of the cutoff\footnote{Let $f(k)$ be a cutoff function that goes to 0 for $k\epsilon\to \infty$, and goes to 1 for $k\epsilon<1$, where $\epsilon$ is a small parameter that has a dimension of time. Let $F(k)$ be the indefinite integral of a function $-2\sin{2k(s-w)}$. Clearly, $F(k_{c})=-1/(s-w)+\mathcal{O}(k_{c}T)$ and $F(\infty)\to0$ as $\epsilon\to0$. Thus, Eq.~(\ref{eq:kinteg1}) does not depend on the cutoff.}. 
With Eqs.~(\ref{eq:kinteg1}) and (\ref{eq:kinteg2}), we obtain 
\begin{align}
    \label{eq:<j>largek}
    &\frac{q}{2\pi}\left(\int^{\infty}_{k_{c}}+\int^{-k_{c}}_{-\infty}\right)dk \left[|h^{I}_{k}|^{2}-|h^{II}_{k}|^{2}+\frac{k}{\omega}\right]\notag
    \\
    &\hspace{0.5cm}=
     \frac{2qm^{2}}{\pi}\int_{t_{0}}^{t}ds
        \left[-\int^{s}_{t_{0}}\frac{\sin{2\Psi(s,w)}}{s-w}dw+\log{|2k_{c}e^{\gamma}(s-t_{0})|}\sin{2\Psi(s,t_{0})}
       \right]\notag
       \\
       &\hspace{5cm}+\mathcal{O}(k_{c}m^{2}T^{2})+\mathcal{O}\left(\frac{m^{4}}{k_{c}^{3}}\right).
\end{align}
Similarly, for the $k\sim m$ region, we first start with noting that 
\begin{align}
     1&=|h^{I}_{k}|^{2}+|h^{II}_{k}|^{2}\notag
    \\
    &=|X^{(0)}|^{2}+|Y^{(0)}|^{2}\notag
    \\
    &+2\mathrm{Re}\left(X^{(0)*}X^{(1)}+Y^{(0)*}Y^{(1)}\right)+\mathcal{O}(m^{2}T^{2}).
\end{align}
Since $|X^{(0)}|^{2}+|Y^{(0)}|^{2}=1$ using Eqs.~(\ref{eq:x0solsmallk}) and (\ref{eq:y0solsmallk}), we have the following constraint
\begin{align}
    \label{eq:xnorm1stsmallk}
    \mathrm{Re}\left(X^{(0)*}X^{(1)}\right)+
    \mathrm{Re}\left(Y^{(0)*}Y^{(1)}\right)=0.
\end{align}
Using Eqs.~(\ref{eq:x0solsmallk}), (\ref{eq:y1solsmallk}), and (\ref{eq:xnorm1stsmallk}), we obtain 
\begin{align}
    \frac{q}{2\pi}\int^{k_{c}}_{-k_{c}}dk \left[|h^{I}_{k}|^{2}-|h^{II}_{k}|^{2}+\frac{k}{\omega}\right]=
    \frac{2q}{\pi}\int^{k_{c}}_{-k_{c}}dk \left[\mathrm{Re}\left(X^{(0)*}X^{(1)}\right)+\mathcal{O}(m^{2}\mathcal{T}^{2})\right].
\end{align}
Then, plugging Eqs.~(\ref{eq:x0solsmallk}) and (\ref{eq:x1solsmallk}) into the above expression yields
\begin{align}
    \label{eq:<j>smallk}
    =&\frac{2q}{\pi}\int^{t}_{t_{0}}ds\int^{k_{c}}_{-k_{c}}dk \mathrm{Re}\left[ik\frac{\omega-k}{2\omega}-i\frac{m^{2}}{2\omega}e^{-2i\Psi(s,t_{0})}\right]+\mathcal{O}(k_{c}m^{2}\mathcal{T}^{2})\notag,
    \\
    =&-\frac{2qm^{2}}{\pi}\int^{t}_{t_{0}}ds\sin{2\Psi(s,t_{0})}\int^{k_{c}}_{-k_{c}}\frac{1}{2\omega}dk +\mathcal{O}(k_{c}m^{2}\mathcal{T}^{2})\notag,
    \\
    =&-\frac{2qm^{2}}{\pi}\int^{t}_{t_{0}}ds\sin{2\Psi(s,t_{0})}\left[
    \log{\left(\frac{2k_{c}}{m}\right)}+\mathcal{O}\left(\frac{m}{k_{c}}\right)
    \right]+\mathcal{O}(k_{c}m^{2}\mathcal{T}^{2}).
\end{align}
Adding Eqs.~(\ref{eq:<j>largek}) and (\ref{eq:<j>smallk}) results in the cancellation of the cutoff dependence,
\begin{align}
    \label{eq:<j>correction}
     &\frac{q}{2\pi}\int^{\infty}_{-\infty}dk \left[|h^{I}_{k}|^{2}-|h^{II}_{k}|^{2}+\frac{k}{\omega}\right]\notag
     \\
     &\qquad=
     \frac{2qm^{2}}{\pi}\int_{t_{0}}^{t}ds
        \left[-\int^{s}_{t_{0}}\frac{\sin{2\Psi(s,w)}}{s-w}dw+\log{|me^{\gamma}(s-t_{0})|}\sin{2\Psi(s,t_{0})}
        +\mathcal{O}
       \right]
\end{align}
where $\mathcal{O}$ denotes the order of correction to the expression in the parentheses and is explicitly given by 
\begin{align}
    \mathcal{O}=\mathcal{O}\left(\frac{m}{k_{c}}\right)+\mathcal{O}(k_{c}\mathcal{T})+\mathcal{O}\left(\frac{m^{2}}{k_{c}^{3}\mathcal{T}}\right)
\end{align}
By choosing $k_{c}=(m/\mathcal{T}^{p-1})^{1/p}$ for an integer $p>1$, the correction terms become
\begin{align}
    \mathcal{O}\left(\frac{m}{k_{c}}\right)\sim&\mathcal{O}\left((m\mathcal{T})^{1-\frac{1}{p}}\right),
    \\
    \mathcal{O}\left(k_{c}\mathcal{T}\right)\sim&\mathcal{O}\left((m\mathcal{T})^{\frac{1}{p}}\right),
    \\
    \mathcal{O}\left(\frac{m^{2}}{k_{c}^{3}\mathcal{T}}\right)\sim&\mathcal{O}\left((m\mathcal{T})^{2-\frac{3}{p}}\right).
\end{align}
Since $p>1$, they are the subleading order terms in the integrand of Eq.~\eqref{eq:<j>correction} in the limit of $m\mathcal{T}\to0$. 
Therefore, we can safely drop correction terms from Eq.~(\ref{eq:<j>correction}). 
Adding $-q^{2}A_{C}(t)/\pi$ to Eq.~(\ref{eq:<j>correction}) gives the expression for $\braket{j^{x}}$, which is Eq.~(\ref{eq:<j>fermismallm}) that we presented in section \ref{sec:fermi Schwinger}.

\bibliography{ScalarField}
\bibliographystyle{utphys}

\end{document}